\documentclass[11pt,letterpaper]{article}

\usepackage[margin=1in]{geometry}
\usepackage{graphicx} % Required for inserting images
\usepackage{amsmath, amssymb, amsthm, bbm, dsfont, bm}
\usepackage[ruled,vlined,algo2e]{algorithm2e}
\usepackage{diagbox}
\usepackage{tikz}
\usepackage{algorithm}
\usepackage{algpseudocode}
\usepackage{amsmath,amssymb}
\usepackage[colorlinks=true,linkcolor=blue,citecolor=blue,urlcolor=blue]{hyperref}
\usepackage{thmtools}
\usepackage{cleveref}
\usepackage{authblk}
\usepackage{fluff}

\usepackage{natbib}
\usetikzlibrary{arrows.meta,positioning,calc,fit,backgrounds,shapes.geometric}

\newcommand{\cconv}{\operatorname{conv}}

\newcommand{\NE}{\mathrm{NE}}

\newcommand{\CCE}{\mathrm{CCE}}

\newcommand{\learners}{\mathcal{L}}
\newcommand{\alg}{\mathcal{A}}
\newcommand{\menu}{\mathcal{M}}
\newcommand{\gameL}{G_{\learners}}

\newcommand{\E}{\mathbb{E}}

\newcommand{\sw}{\mathrm{sw}}

\usepackage{xcolor}

\title{Non-manipulability and Approachability Beyond Two-Player Games}
\author[1]{Natalie Collina}
\author[2]{Elijah Journey Fullerton}
\author[2]{Jingyi Liu}
\author[2]{S. Matthew Weinberg}

\affil[1]{Massachusetts Institute of Technology}
\affil[2]{Princeton University}

\date{}

\begin{document}

\maketitle

\begin{abstract}
    We study the problem of guaranteeing non-manipulability and low regret for $n-1$ independent learners in an $n$-player repeated general game with one optimizer. Non-manipulability is a property which guarantees that learning agents can prevent a strategic opponent from exploiting their learning dynamics for its own benefit, while no-regret provides guarantees to a learner with respect to a fixed static benchmark. This problem is well understood in games with a single learner and a single optimizer, where the intersection of non-manipulable and no-regret algorithms is exactly the class of all no-swap regret algorithms. But having multiple independent learners introduces a fundamental constraint: without shared randomness, the learners' joint mixed actions must remain product distributions and individual no-swap regret guarantees need not provide a guarantee of non-manipulability. Our core contributions are as follows.
    \begin{itemize}
    \item We introduce a set of $n-1$ algorithms that $n-1$ learners can play which individually guarantee sublinear swap regret and, when played in conjunction with each other, jointly guarantee non-manipulability against an optimizer.    
    \item These algorithms require the computation of a Nash equilibrium at every timestep, and we show that this computational cost is inherent in general games. We construct a reduction from computing Nash in $(n-1)$-player games to finding $n-1$ jointly non-manipulable no-external regret algorithms. Therefore, polynomial time algorithms which guarantee multi-learner non-manipulability in conjunction with no-external regret do not exist unless PPAD $\subseteq$ FBPP.
    \item In order to construct these algorithms, we develop new techniques for Blackwell approachability beyond $2$-player games. We provide a general characterization for sets which are approachable by $n-1$ players without correlated randomness. 
    \end{itemize}

\end{abstract}

\newpage
\section{Introduction}
Consider an online marketplace in which $n$ sellers deploy algorithms that price and sell goods over time. Each seller faces a complex task in determining which algorithm to select; they must be competitive with the other sellers, but they also must reckon with potential manipulation from the marketplace owner. The marketplace owner has more information than any individual seller---they can collect large amounts of data about the behavior of the algorithms on the platform and design rules to skim as much profit off of them as possible. Meanwhile, the sellers may not even know the owner's utility function. That is, even if the sellers generally believe that the owner seeks to maximize their own profit, the sellers may not know exactly how the owner derives their profits. Thus, sellers may want to use algorithms which not only provide long-term strategic guarantees with respect to each other, such as regret guarantees, but which level the strategic playing field with respect to the marketplace owner. In other words, sellers may find it desirable to deploy algorithms which induce strategic responses which are \emph{simple} to reason about even with limited information.

In online learning, the notion of \emph{non-manipulability} captures this simplicity. In two-player repeated games, an algorithm is non-manipulable if it is always a best response of the opponent to play the same static strategy across all rounds. In this paper, we ask: in an $n$-player game, can we construct $n-1$ algorithms which each attain no regret and, when played together, induce a static best response for the marketplace owner? We call this notion \emph{joint} non-manipulability, and we are interested in understanding when it can be attained by a set of algorithms which also attain no-regret. This question is well understood in two-player settings: the set of no-regret non-manipulable algorithms is exactly the set of no-swap regret algorithms. When moving to $n$-player games, there is a natural reduction to the $2$-player case if the sellers employ shared private randomness. But there are many reasons why sharing randomness is infeasible; there may be a lack of trust between the sellers, and information sharing between competitors in marketplaces is often tightly regulated.% so the action may be illegal. 

Without shared private randomness, moving to the $n$-player setting fundamentally changes the landscape of no-regret and non-manipulability. If every seller independently plays a no-swap regret algorithm, this does not guarantee that they will be jointly non-manipulable (Lemma~\ref{lem:BMisManipulable}). However, in our first main result, we show that it is still possible to construct $n-1$ algorithms which have vanishing regret and are jointly non-manipulable (Theorem~\ref{thm:main-NM-algo}). The algorithms we construct require computing a Nash equilibrium of an auxiliary game at every round. In our second main result, we show that this computational difficulty is inherent: we give a reduction from computing Nash equilibria to finding jointly non-manipulable no-external regret algorithms, establishing that polynomial time no-regret and jointly non-manipulable algorithms do not exist unless PPAD $\subseteq$ FBPP (Theorem~\ref{thm:universal-nonmanip-hardness}). Finally, we provide a general characterization of which convex sets can be jointly approached by multiple learners without shared randomness, extending the classical approachability framework to this setting (Theorem~\ref{thm:joint-approachability-characterization}).

\subsection{Context \& Related Work}

\paragraph{Non-manipulability.} The concept of non-manipulability has (at least) two interpretations in online learning. The first is quantitative: when viewed as a Stackelberg game between a single learner and a single optimizer, non-manipulable algorithms allow the learner to achieve their Stackelberg value on average per round~(\cite{DengSchneiderSivan2019}). The second is qualitative: simplicity is a common-sense desideratum in strategic interactions. The value of simplicity has been remarked upon in many contexts; see, for example, \cite{Wilson1987} %\nrc{not sure how to format this bc I don't want to do nested parentheses} 
and the long line of work on simple vs.~optimal auctions following~\cite{HartlineR09}. It is \emph{much} simpler for a learner to engage with a static game every round than with an optimizer who dynamically manipulates them. 

Our work takes the qualitative motivation, and posits that non-manipulability is desirable \emph{for the online learners} because it relieves \emph{the optimizer} of the need to dynamically optimize, thereby simplifying the ultimate interaction \emph{for the online learners} and making the entire process more robust. One can draw the following analogy to simple vs.~optimal auctions: simple auctions are desirable \emph{for the auctioneer} because they relieve \emph{the bidders} of the need to understand complex auctions and optimize bids, thereby simplifying the ultimate interaction \emph{for the auctioneer} and making the entire process more robust and predictable. Simplicity in our motivating example is desirable especially because the platform owner has an informational advantage over the sellers, including its own payoff function, and therefore any attempt by the sellers to predict the owner's dynamic behavior is error-prone.

\paragraph{Non-manipulability for multiple learners.} There is a strong literature following~\cite{BravermanMaoSchneiderWeinberg2018} and~\cite{DengSchneiderSivan2019} on non-manipulability for a single learner. However, consideration of multiple online learners has proven surprisingly stubborn. On this front, the only progress concerns repeated online auctions~(\cite{CaiWeinbergWildenhainZhang2023, KumarSchneiderSivan2024, Zhao2026, CaiLuoWeiZheng2026}). Here, the optimizer's payoff is known because it is the expected revenue of the auction and the game is a repeated single-item auction, so ``non-manipulability'' in this setting means ``the optimizer cannot improve their expected revenue beyond running Myerson's optimal auction~(\cite{Myerson81}) each round,'' and these claims crucially leverage both the structure of the game and the optimizer's payoff. In contrast, our work is the first to develop non-manipulable online learning algorithms for multiple learners in arbitrary games and, in fact, any games beyond single-item auctions. 

\paragraph{Joint online learning algorithms.} Our online learning algorithms provide two guarantees. First, they guarantee no-regret \emph{no matter what any other player does}. Second, they guarantee joint non-manipulability \emph{if all learners successfully coordinate}. This contributes to a rich literature of similar guarantees. For example, ~\cite{daskalakis2011near}
give algorithms that guarantee no regret against arbitrary opponents
while achieving faster convergence to equilibrium when both players
of a zero-sum game follow the prescribed dynamics.
Subsequent work develops analogous guarantees in general-sum games
for external regret (\cite{syrgkanis2015fast,daskalakis2021near})
and for internal and swap
regret (\cite{anagnostides2022near}).

\paragraph{Approachability.} Consider a learner and an optimizer playing a repeated vector-payoff game, i.e., each player's payoff is a vector rather than a scalar. Approachability asks for a given closed, convex set $S$ whether the learner can force the time-averaged payoff arbitrarily close to $S$ over time, and Blackwell's approachability theorem provides two equivalent necessary and sufficient conditions~(\cite{Blackwell1956}). The first is response satisfiability: is it the case that, for every possible action $a$ of the optimizer, the learner has a response $b$ such that the payoff vector of $(a,b)$ lies in $S$? The second is halfspace satisfiability: for all halfspaces $H$ containing $S$, does there exist a learner strategy $b_H$ such that for all actions $a$ of the optimizer, the payoff vector of $(a,b_H)$ lies in $H$? Despite being equivalent conditions for two-player games, the natural generalizations of these conditions are \emph{not} equivalent with multiple learners. Indeed, a natural generalization of halfspace satisfiability turns out to remain necessary and sufficient, while response satisfiability is neither sufficient nor necessary. See Section~\ref{sec:characterization-of-approachable-sets} for further details.
\newpage
\paragraph{Key Takeaways.} We design online learning algorithms that: (a) guarantee no-regret no matter what other players do, and (b) guarantee joint non-manipulability whenever all learners coordinate. Using these algorithms removes the incentive for a platform owner to dynamically optimize, resulting in a simpler and more robust/predictable experience for the learners. Moreover, this is the first jointly non-manipulable algorithm beyond single-item auctions, and unlike the work in auctions does not make assumptions on the structure of the game or knowledge of the optimizer's particular objective. Our work additionally contributes to a long line of work considering learning algorithms with both worst-case guarantees and additional guarantees in case of coordination. Finally, our results follow from an extension of Blackwell approachability to multiple learners, which identify halfspace satisfiability as the ``right'' necessary and sufficient condition to extend, even though it is equivalent to response satisfiability with a single learner.

Finally, our algorithms are information theoretic, and require computing a Nash equilibrium. We further show that such computation is necessary via reduction from $(n-1)$-player Nash to $(n-1)$-learner non-manipulability.

For a more extensive treatment of related work, see Appendix$~\ref{sec:ext-related-work}$.

\section{Preliminaries}
\label{sec:preliminaries}

\paragraph{Game and repeated-play model.}
Let
\(
    G=(A_0,A_1,\ldots,A_{n-1},u_0,u_1,\ldots,u_{n-1})
\)
be a finite normal-form game, where player $0$ is the \emph{optimizer} and
$\learners=\{1,\ldots,n-1\}$ are the \emph{learners}. $A_i$ and $u_i$ are, respectively, the action set and the utility function of player $i$. Let
\[
    A_{\learners}:=\prod_{i\in\learners}A_i,
    \qquad
    A:=A_0\times A_{\learners},
\]
and assume $u_i:A\to[-1,1]$ for every player $i$.
For a finite set $S$, let $\Delta(S)$ denote the simplex over $S$, and let
$e_a$ denote the point mass on $a\in S$. We extend utilities multilinearly
to distributions over action profiles.

The game is repeated for $T$ rounds. In round $t$, learner $i$ selects
$x_i^{(t)}\in\Delta(A_i)$. Since the learners randomize independently, their
joint mixed action is the product distribution
\[
    q^{(t)}
    :=
    \bigotimes_{i\in\learners}x_i^{(t)}
    \in\Delta(A_{\learners}).
\]
The optimizer observes $q^{(t)}$ and then chooses
$y^{(t)}\in\Delta(A_0)$ and draws $c^{(t)}\sim y^{(t)}$.
The learners subsequently draw their actions independently from
$q^{(t)}$. The optimizer may choose $y^{(t)}$ adaptively as a function of
the history and the learners' current product distribution.

We assume throughout that the learners know the action sets and learner
utilities $u_1,\ldots,u_{n-1}$, but need not know the optimizer utility $u_0$.
For our menu-based analysis, we consider learner algorithms whose mixed
strategies are determined by the optimizer's previously observed actions
$c^{(1)},\ldots,c^{(t-1)}$. In particular, the learners' private action
samples do not affect their subsequent mixed strategies, and any
algorithmic tie-breaking is fixed in advance. We separately
consider an extension that uses realized payoff feedback to detect learner
defections.
\newpage
\paragraph{Statistics of Repeated Play.}
Two key metrics that will be relevant throughout this paper are
\emph{external regret} and its refinement, \emph{swap regret}.

\begin{definition}[External Regret]
For learner $i$, define the cumulative external regret
\[
    R_i^{\mathrm{ext}}(T)
    :=
    \max_{b_i\in A_i}
    \sum_{t=1}^T
    \left[
        u_i\!\left(c^{(t)},b_i,a_{-i}^{(t)}\right)
        -
        u_i\!\left(c^{(t)},a_i^{(t)},a_{-i}^{(t)}\right)
    \right].
\]
In other words, external regret (sometimes shortened to \emph{regret}) is the gap between the current utility and the best single action, keeping history fixed. An algorithm has \emph{no-external regret} if its expected external
regret is $o(T)$ uniformly over all possible opponent policies.
\end{definition}

Swap regret is a refinement of external regret which provides a stronger
benchmark. Instead of competing only with the best fixed action in
hindsight, swap regret compares the learner's realized sequence of actions
against every fixed remapping of those actions. In particular, when swap regret is sublinear, replacing any fixed
action $a$ by a fixed alternative on the rounds when $a$ was played
cannot improve cumulative payoff by a linear-in-$T$ amount.

\begin{definition}[Swap Regret]
For learner $i$ and a swap
$\phi_i:A_i\to A_i$, define the cumulative swap gain
\[
    R_{i,\phi_i}(T)
    :=
    \sum_{t=1}^T
    \left[
        u_i\!\left(
            c^{(t)},
            \phi_i(a_i^{(t)}),
            a_{-i}^{(t)}
        \right)
        -
        u_i\!\left(
            c^{(t)},
            a_i^{(t)},
            a_{-i}^{(t)}
        \right)
    \right].
\]
The swap regret of learner $i$ is
\[
    R_i^{\mathrm{sw}}(T)
    :=
    \max_{\phi_i:A_i\to A_i}R_{i,\phi_i}(T).
\]
An algorithm has \emph{no swap regret} if its expected swap regret
is $o(T)$ uniformly over the relevant opponent policies.
\end{definition}

These quantities measure regret on a particular realized sequence of
play. A no-regret \emph{algorithm} guarantees sublinear expected regret
against the relevant class of opponent policies. We will also express
regret as a property of distributions over joint actions.

In a finite normal-form game, a player's utility and the gain from any
fixed external or swap deviation depend only on the empirical
distribution over joint actions, not on their order. Thus, while one
could evaluate the utility, external regret, and swap regret of any
player by looking at the full transcript of play, it is sufficient to
look just at this distribution. Following~\cite{ArunachaleswaranCollinaSchneider2024}, we call this distribution over joint
actions a \emph{correlated strategy profile}.

\begin{definition}[Correlated Strategy Profile]
A \emph{correlated strategy profile} (CSP) is a distribution
$\mu\in\Delta(A)$. The realized empirical CSP of a $T$-round
transcript is
\[
    \widehat\mu_T
    :=\frac1T\sum_{t=1}^T
    e_{(c^{(t)},a_1^{(t)},\ldots,a_{n-1}^{(t)})}.
\]
For a profile of learner algorithms $\alg_\learners$ and an optimizer
policy $\pi\in \Pi$, where $\Pi$ is the set of optimizer policies, let
\[
    \varphi_T(\pi,\alg_\learners)
    :=
    \E\left[
        \frac1T\sum_{t=1}^T
        e_{c^{(t)}}\otimes q^{(t)}
    \right]
    \in\Delta(A)
\]
denote the expected empirical CSP induced after $T$ rounds.
\end{definition}

A distribution over repeated rounds of play need not be a product
distribution, even when the algorithms do not share randomness. Consider,
for example, two learners who both play action 1 on even rounds and
action 2 on odd rounds.

We will also use the \emph{swap-regret inequalities} of a CSP: a CSP
$\mu\in\Delta(A)$ satisfies learner $i$'s swap-regret inequalities if
\[
    \sum_{a\in A}\mu(a)
    \left[u_i\!\left(a_0,\phi_i(a_i),a_{-i}\right)-u_i(a)\right]
    \le 0
    \qquad \text{for every }\phi_i:A_i\to A_i.
\]
External regret corresponds to the special case of constant swap functions $\phi_i(a_i)=b_i$. We will use these inequalities to characterize CSPs with no swap regret. In particular, if every CSP
in the asymptotic menu of our learning algorithms satisfies these inequalities, then the algorithms guarantee no swap regret.

In this work it will be useful to think about the set of CSPs that the
optimizer can induce against the learner algorithms. We call this the
\emph{menu} of the learners.

\begin{definition}[Learner Menu]
For a profile of learner algorithms $\alg_\learners$ whose mixed
strategies are determined by the optimizer's past realized actions,
the horizon-$T$ menu of the learners is
\[
    \menu_T(\alg_\learners)
    :=
    \operatorname{conv}
    \left\{
        \frac1T\sum_{t=1}^T e_{c^{(t)}}\otimes q^{(t)}:
        (c^{(1)},\ldots,c^{(T)})\in A_0^T
    \right\},
\]
where for each optimizer-action sequence, $q^{(t)}$ is the product
distribution prescribed by $\alg_\learners$ after observing
$c^{(1)},\ldots,c^{(t-1)}$. We write $\menu(\alg_\learners)$ for the
limit of $\menu_T(\alg_\learners)$ in Hausdorff distance when it exists.
\end{definition}

This notion extends the definition of menus introduced in
\cite{ArunachaleswaranCollinaSchneider2024} to capture dynamics beyond
two-player games. In the original paper, a menu is defined with respect
to a single learner algorithm, and represents the convex set of CSPs that
the optimizer can induce against it. Here, we define the learner menu
with respect to all $n-1$ learner algorithms, and it again represents the
convex set of CSPs that the optimizer can induce against them. Equivalently, the horizon-\(T\) menu is the set of expected CSPs that the optimizer can induce against the learners, ranging over all optimizer policies. This equivalence holds because the learners' mixed strategies depend only on the optimizer's past realized actions. Thus, each optimizer-action sequence determines a unique average CSP, and the set of expected CSPs induced by optimizer policies is exactly the convex hull of these points.

We next define the set of CSPs that will serve as the target of our construction, formed by taking the convex hull of outcomes in which the learners play a Nash equilibrium in response to a fixed optimizer strategy.

\begin{definition}[Nash Response Target]
Let
\(
    \gameL
    :=
    (A_0,A_1,\ldots,A_{n-1},u_1,\ldots,u_{n-1})
\)
denote the game information available to the learners.
For $y\in\Delta(A_0)$, let $\gameL(y)$ denote the induced
$(n-1)$-player game among the learners, with utilities
\[
    u_i^y(a_1,\ldots,a_{n-1})
    :=
    \E_{c\sim y}
    \left[
        u_i(c,a_1,\ldots,a_{n-1})
    \right].
\]
Let $\NE(\gameL(y))\subseteq\Delta(A_{\learners})$ denote the set of
product mixed Nash equilibria of this induced game. Define
\[
    S_{\NE}(\gameL)
    :=
    \operatorname{conv}
    \left\{
        y\otimes q:
        y\in\Delta(A_0),\
        q\in\NE(\gameL(y))
    \right\}.
\]
Every point of $S_{\NE}(\gameL)$ satisfies the individual swap-regret
inequalities of every learner.    
\end{definition}

We now formalize our central strategic guarantee: that the optimizer cannot benefit from adapting its strategy over time rather than committing to a fixed mixed action.

\begin{definition}[Non-manipulability]
A profile of learner algorithms $\alg_\learners$ is
\emph{jointly non-manipulable} if, for every optimizer utility $u_0$,
the optimizer obtains asymptotically no more utility by adapting dynamically
than by committing to a fixed mixed action:
\[
    \sup_{\pi^T\in\Pi^T}
    \left\langle
        u_0,
        \varphi_T(\pi^T,\alg_\learners)
    \right\rangle
    \le
    \sup_{y\in\Delta(A_0)}
    \left\langle
        u_0,
        \varphi_T(y^{\otimes T},\alg_\learners)
    \right\rangle
    +o(1).
\]
\end{definition}

\section{Jointly Non-Manipulable and No-Swap-Regret Algorithms}
\label{sec:algorithm}

We now construct a profile of learning algorithms for the $n-1$ learners that ensure both individual no-swap regret and joint non-manipulability. Throughout this section, we consider the coordinated model in which
every learner follows the prescribed algorithm. We later show in
\Cref{app:learner-defection} how to augment the construction so that each
learner individually retains no swap regret if another learner defects from
the coordinated protocol.

A key conceptual step in understanding our algorithm is the insight that the properties which we are interested in guaranteeing, low-regret and non-manipulability, can be encoded as geometric constraints in correlated strategy profile (CSP) space. 
For every sequence of optimizer actions, the learners' algorithms induce an empirical CSP representing the average distribution of joint play. The learners' menu is the convex hull of all such CSPs.

Recall that each CSP $\varphi(\pi, \alg_\learners)$ is a vector in $\Delta(A)$ which denotes the long-term average distribution over $n$ player's joint action space, when the learners commit to some learning algorithms $\alg_\learners$ and the optimizer uses some policy $\pi$. The learners' menu $\menu(\alg_\learners)$ is then the convex hull of all such points that can be induced by any optimizer policy $\pi\in \Pi$. The problem of constructing non-manipulable and no-swap-regret algorithms then reduces to the problem of constructing algorithms whose menu $\menu\subseteq\Delta(A)$ satisfy the properties described in the following lemma.

\begin{lemma}
\label{lem:menu-criterion}
Fix a finite game with utilities in $[-1,1]$, and let
$\alg_\learners$ be a profile of learner algorithms with
asymptotic menu $\menu=\menu(\alg_\learners)$.
For learner $i$, let $\Phi_i:=\{\phi_i:A_i\to A_i\}$ and define
\[
G_{i,\phi_i}(\mu):=
\sum_{a\in A}\mu(a)
\bigl[u_i(a_0,\phi_i(a_i),a_{-i})-u_i(a)\bigr].
\]
Suppose that:
\begin{enumerate}
\item Every $\mu\in\menu$ has nonpositive swap gain for every learner:
$G_{i,\phi_i}(\mu)\le 0$ for all $i\in\learners$ and
$\phi_i\in\Phi_i$.
\item Every extreme point $s\in\menu$ is a product distribution
$s=y_s\otimes q_s$, and is asymptotically attainable by a static
optimizer strategy, in the sense that
\[
 \inf_{y\in\Delta(A_0)}
 \left\|\varphi_T(y^{\otimes T},\alg_\learners)-s\right\|_1=o(1).
\]
\end{enumerate}
Then every learner has no swap regret, and the profile is jointly
non-manipulable.
\end{lemma}

We defer the proof of this lemma to Appendix$~\ref{app:menu-criterion-proof}$

\paragraph{No-regret non-manipulable menu.} We now define a desired set that satisfies both properties. For an optimizer mixed action
\(y\in\Delta(A_0)\), let \(G_\learners(y)\) denote the game induced among the
\(n-1\) learners when the optimizer plays \(y\). Let
$\NE(\gameL(y))$ denote the set of mixed Nash equilibria of this
induced game. Define the Nash-response target set
\[
    S_{\NE}(\gameL)
    =
    \operatorname{conv}
    \left\{
        y\otimes q:
        y\in\Delta(A_0),\
        q\in\NE(\gameL(y))
    \right\}.
\]

Every extreme point of $ S_{\NE}(\gameL)$ is a product distribution by definition. Every extreme point also guarantees no-swap-regret for each learner: if the
optimizer plays \(y\) and the learners play a Nash equilibrium of $G_\learners(y)$,
then no learner benefits from any unilateral swap deviation. By linearity, the
same individual swap inequalities hold throughout $ S_{\NE}(\gameL)$. 

\paragraph{Approachability algorithm.} What remains to be shown is that there exists a set of independent learning algorithms $\alg_\learners$ whose menu is $ S_{\NE}(\gameL)$. Our construction has two components. We first give an approachability procedure in \Cref{alg:joint-non-manipulable-algorithm} that prevents any optimizer from inducing an
outcome outside $S_{\NE}(\gameL)$. We then apply the menu-completion
construction of \Cref{lem:menu-completion}, which makes every Nash-response
outcome attainable by an optimizer. Together these imply
that the learners' asymptotic menu is exactly $S_{\NE}(\gameL)$.

% The goal in round $t$ is to choose an implementable product distribution $q^{(t)}$ and a new target point $s^{(t)}\in S_\NE$ such that, regardless of the optimizer's action, the new discrepancy has non-positive inner product with $U^{(t)}$, which is a necessary and sufficient condition for approaching $S_\NE$ (see \Cref{sec:characterization-of-approachable-sets}).

In each round $t$, our algorithm compares the action $k^{(t)}=e_{c^{(t)}}\otimes q^{(t)}\in\Delta(A)$ generated by actual play with a target point $s^{(t)}\in S_{\NE}$. The algorithm maintains the cumulative discrepancy between the empirical distributions of play and the sequence of target points constructed so far as 
$$U^{(t)}:=\sum_{\tau=1}^{t-1}\left(k^{(\tau)}-s^{(\tau)}\right)\in\mathbb{R}^{|A|}.$$

Given $U^{(t)}$, define the auxiliary game
\[
    \widetilde G^{(t)}
    =
    \left(
        A_0,\ldots,A_{n-1},
        \widetilde u_0^{(t)},
        u_1,\ldots,u_{n-1}
    \right),
\]
where the learners retain their original utilities and the optimizer's
auxiliary utility is
\[
    \widetilde u_0^{(t)}
    \left(
        y,x_1,\ldots,x_{n-1}
    \right)
    :=
    \left\langle
        U^{(t)},
        y\otimes
        x^{(t)}_1\otimes\cdots\otimes x^{(t)}_{n-1}
    \right\rangle.
\]

\begin{algorithm2e}[t]
\caption{Jointly non-manipulable and no-swap regret algorithms}
\label{alg:joint-non-manipulable-algorithm}
\DontPrintSemicolon

\KwIn{
$\gameL=(A_0,\ldots,A_{n-1},u_1,\ldots,u_{n-1})$
}

Initialize $U^{(1)}\leftarrow 0$\;

\For{$t=1,\ldots,T$}{
    Construct the auxiliary game $\widetilde G^{(t)}$\;

    Compute a mixed Nash equilibrium
    \(
        \left(
            \widetilde y^{(t)},
            \widetilde x_1^{(t)},
            \ldots,
            \widetilde x_{n-1}^{(t)}
        \right)
    \)
    of $\widetilde G^{(t)}$\;

    Set
    \(
        q^{(t)}
        :=
        \bigotimes_{i\in\learners}
        \widetilde x_i^{(t)};
    \)

    Each learner $i$ plays
    $x_i^{(t)}=\widetilde x_i^{(t)}$\;

    The optimizer observes $q^{(t)}$, chooses
    $y^{(t)}\in\Delta(A_0)$, and samples
    $c^{(t)}\sim y^{(t)}$\;

    Each learner independently samples
    $a_i^{(t)}\sim x_i^{(t)}$\;

    Set
    \(
        k^{(t)}
        :=
        e_{c^{(t)}}\otimes q^{(t)}
    \)
    and
    \(
        s^{(t)}
        :=
        \widetilde y^{(t)}\otimes q^{(t)};
    \)

    Update
    \(
        U^{(t+1)}
        :=
        U^{(t)}+k^{(t)}-s^{(t)}.
    \)
}
\end{algorithm2e}

A Nash equilibrium $\widetilde y^{(t)}\otimes q^{(t)}$ of this auxiliary game simultaneously provides the two objects required for the approachability update: the joint action of the learners $q^{(t)}$ and the target point $s^{(t)}=\widetilde y^{(t)}\otimes q^{(t)}$.

First, note that because the learners are mutual best responses under their original utilities,
\(
    q^{(t)}
    \in
    \NE\!\left(
        \gameL(\widetilde y^{(t)})
    \right),
\)
and hence
\(
    s^{(t)}
    =
    \widetilde y^{(t)}\otimes q^{(t)}
    \in
    S_{\NE}(\gameL).
\)
Second, because $\widetilde y^{(t)}$ is a best response in the auxiliary
game,
\[
    \left\langle
        U^{(t)},
        e_c\otimes q^{(t)}
    \right\rangle
    \le
    \left\langle
        U^{(t)},
        \widetilde y^{(t)}\otimes q^{(t)}
    \right\rangle
    \qquad
    \forall c\in A_0.
\]
Taking $c=c^{(t)}$ gives
\begin{equation}
\label{eq:directional-inequality}
    \left\langle
        U^{(t)},
        k^{(t)}-s^{(t)}
    \right\rangle
    \le 0.
\end{equation}

The following lemma shows that these properties imply that the realized distribution of play approaches the target set $S_\NE$.

\begin{lemma}
\label{lem:product-play-approachability}
For every optimizer policy and every realized path of optimizer actions, Algorithm~\ref{alg:joint-non-manipulable-algorithm} satisfies
\[
    \operatorname{dist}
    \left(
        \frac1T\sum_{t=1}^T k^{(t)},
        S_{\NE}(\gameL)
    \right)
    \le
    \sqrt{\frac{2}{T}}.
\]

By the convexity of $S_{\NE}(\gameL)$ and the definition of the
horizon-$T$ menu, this also implies
\[
 \sup_{\mu\in\menu_T(\alg_\learners)}
 \operatorname{dist}(\mu,S_{\NE}(\gameL))
 \le \sqrt{\frac{2}{T}}.
\]
In particular, any asymptotic menu of the base algorithm is contained
in $S_{\NE}(\gameL)$.
\end{lemma}

\begin{proof}
    Since
\(
    U^{(t+1)}
    =
    U^{(t)}+k^{(t)}-s^{(t)},
\)

we have
\begin{align*}
    \|U^{(t+1)}\|_2^2
    &=
    \|U^{(t)}\|_2^2
    +
    2
    \left\langle
        U^{(t)},
        k^{(t)}-s^{(t)}
    \right\rangle
    +
    \|k^{(t)}-s^{(t)}\|_2^2\\
    &\le
    \|U^{(t)}\|_2^2
    +
    \|k^{(t)}-s^{(t)}\|_2^2,
\end{align*}
where the inequality follows from
\eqref{eq:directional-inequality}.
Both $k^{(t)}$ and $s^{(t)}$ are probability distributions over $A$.
Therefore
\(
    \|k^{(t)}-s^{(t)}\|_2^2
    \le 2.
\)
Since $U^{(1)}=0$, summing over rounds yields
\(
    \|U^{(T+1)}\|_2
    \le
    \sqrt{2T}.
\)
Note that,
\[
    \frac1T
    \sum_{t=1}^T
    s^{(t)}
    \in
    S_{\NE}(\gameL)
\]
because each $s^{(t)}\in S_{\NE}(\gameL)$ and
$S_{\NE}(\gameL)$ is convex. Hence
\begin{align*}
    \operatorname{dist}
    \left(
        \frac1T\sum_{t=1}^T k^{(t)},
        S_{\NE}(\gameL)
    \right)
    &\le
    \left\|
        \frac1T
        \sum_{t=1}^T
        \left(
            k^{(t)}-s^{(t)}
        \right)
    \right\|_2 =
    \frac{\|U^{(T+1)}\|_2}{T}\le
    \sqrt{\frac2T}.
\end{align*}
\end{proof}

\newpage
Finally, we remark on why our algorithm is jointly non-manipulable. Lemma~\ref{lem:product-play-approachability} guarantees the upper
containment
\(
    \menu(\alg_\learners)
    \subseteq
    S_{\NE}(\gameL).
\)
For joint non-manipulability we additionally want every Nash-response outcome
to be attainable by an appropriate static optimizer. We therefore apply the menu-completion construction of
\Cref{lem:menu-completion}.

We are now ready to state our main theorem.
\begin{theorem}
\label{thm:main-NM-algo}
Let $\alg_\learners$ be the profile in
\Cref{alg:joint-non-manipulable-algorithm}. %Then for every $T$,
%\[
% \sup_{\mu\in\menu_T(\alg_\learners)}
% \operatorname{dist}(\mu,S_{\NE}(\gameL))
% \le \sqrt{\frac{2}{T}}.
%\]
Then we can modify Algorithm 1 so that the resulting algorithms' asymptotic menu $\menu(\mathcal{A}_{\learners}')=S_{\mathrm{NE}}(\gameL)$.
\end{theorem}
\begin{proof}
The proof follows from Lemma~\ref{lem:menu-criterion}, Lemma~\ref{lem:product-play-approachability}, and Lemma~\ref{lem:menu-completion}.
\end{proof}

We prove the analagous theorem which incorporates handling learner defection in Appendix$~\ref{sec:unified-algorithm}$.

\section{Hardness of Universal Non-Manipulability}
\label{sec:hardness}

In the previous section, we showed that multiple independent learners can
jointly obtain low swap regret and non-manipulability through an approachability algorithm. The resulting algorithm, however, requires the computation of
a Nash equilibrium of an auxiliary game in every round. % in order to obtain an implementable mixed strategy and a target point that satisfy the approachability condition in \eqref{eq:directional-inequality}. 
This raises a natural
question: is this equilibrium computation merely an artifact of our
construction, or is it inherent to universal multi-learner
non-manipulability?

We show that the computational difficulty is inherent in general games. We
give a reduction from approximate Nash-equilibrium computation in an arbitrary
bimatrix game to no-swap regret and joint non-manipulability in a three-player game.\footnote{In fact, the reduction easily extends to Nash-equilibrium in $(n-1)$-player games to no-swap regret and joint non-manipulability in a game with $n-1$ learners and one optimizer.} The key idea is
to construct an optimizer utility so that, after observing the learners' current
product distribution $q$, if $q$ is not a learner's Nash equilibrium, then the optimizer can exploit how much a learner can gain by unilaterally deviating from $q$. 
%In contrast, if the optimizer is static, the learners will always play some Nash equilibrium that the optimizer cannot exploit. 
In fact, our hardness result holds even for the strictly weaker requirement that the learners' algorithms are jointly non-manipulable and guarantee no-external regret. %To prevent the optimizer from obtaining an asymptotic payoff above $V_{\CCE}$, the learners should play some distribution $q$ that is a CCE. %Additionally, the constraint that the learners are independent and must play a product distribution conditioned on history narrows $q$ down to a Nash equilibrium.

\paragraph{Hardness menu.} $V_{\CCE}$ is a benchmark for the optimizer's payoff which denotes its maximum payoff when playing a static strategy against learners that respond with a coarse correlated equilibrium (and tie-breaks in favor of the optimizer). More formally,
let
\[
    S_{\CCE}(\gameL)
    =
    \cconv
    \left\{
        y \otimes q
        :
        y \in \Delta(A_0),
        \ q \in \CCE(\gameL(y))
    \right\}
\]
where $\CCE(\gameL(y))$ is the set of coarse correlated equilibrium of the induced game $\gameL(y)$.
Define
\[
V_{\CCE}(\gameL,u_0):=\max_{\mu\in S_{\CCE}(\gameL)}\langle u_0,\mu\rangle,
  \label{eq:VCCE}
\]
Note, for all games, $S_\NE\subseteq S_\CCE$, therefore any algorithm which approaches $S_\NE$ also approaches $S_\CCE$. Additionally, when learners have menu $\menu\subseteq S_\CCE$, the optimizer's asymptotic payoff is at most $V_\CCE$.

% maximum expected payoff is exactly the Nash
% exploitability of that distribution. At the same time, the optimizer's
% correlated-equilibrium response benchmark is zero. Consequently, any universal
% learning algorithm that prevents the optimizer from asymptotically obtaining
% payoff above this benchmark can be used to find an approximate Nash
% equilibrium.

\paragraph{Reduction.} We are now ready to show how to construct a repeated game instance that computes an approximate Nash equilibrium of any bimatrix game $G$. Let
\(
    G=(A,B)
\)
be an arbitrary two-player bimatrix game with finite action sets $A_1$ and
$A_2$. We normalize both players' payoffs to lie in $[0,1]$. For mixed
strategies $x\in\Delta(A_1)$, $z\in\Delta(A_2)$, 
write
\[
    u_1(x,z)=x^\top A z,
    \qquad
    u_2(x,z)=x^\top B z.
\]

Define the Nash exploitability of $(x,z)$ by
\[
    \operatorname{Exploit}_G(x,z)
    :=
    \max
    \left\{
        0,\,
        \max_{i\in A_1}
        \left[
            u_1(i,z)-u_1(x,z)
        \right],\,
        \max_{j\in A_2}
        \left[
            u_2(x,j)-u_2(x,z)
        \right]
    \right\}.
\]
Thus,
$\operatorname{Exploit}_G(x,z)\le\varepsilon$
if and only if $(x,z)$ is an $\varepsilon$-approximate Nash equilibrium of
$G$.

We construct a three-player game $\widehat G$. Players $1$ and $2$ are the
learners and player $0$ is the optimizer. The learners retain their utilities
from $G$, independently of the optimizer's action:
\[
    \widehat u_1(c,a_1,a_2)
    :=
    A_{a_1,a_2},
    \qquad
    \widehat u_2(c,a_1,a_2)
    :=
    B_{a_1,a_2}
\]
for every optimizer action $c$. The optimizer's action set is
\[
    A_0
    :=
    \{0\}
    \cup
    \{R_i:i\in A_1\}
    \cup
    \{C_j:j\in A_2\}.
\]

The null action has payoff
\(
    \widehat u_0(0,a_1,a_2)=0.
\)

For every $i\in A_1$, define
\(
    \widehat u_0(R_i,a_1,a_2)
    :=
    A_{i,a_2}-A_{a_1,a_2},
\)

For every $j\in A_2$, define
\(
    \widehat u_0(C_j,a_1,a_2)
    :=
    B_{a_1,j}-B_{a_1,a_2}.
\)

The optimizer's deviation actions directly measure the learners' gains from
switching to fixed pure actions.

\begin{lemma}
\label{lem:optimizer-exploitability}
If the learners play the product distribution $x\otimes z$, then
\[
    \max_{c\in A_0}
    \E_{(a_1,a_2)\sim x\otimes z}
    \left[
        \widehat u_0(c,a_1,a_2)
    \right]
    =
    \operatorname{Exploit}_G(x,z).
\]
\end{lemma}
We defer the proof of this lemma to Appendix$~\ref{app:optimizer-exploitability}$.

Because the optimizer observes the learners' current product distribution
before choosing its action, in every round it can choose an action attaining
the maximum in Lemma \ref{lem:optimizer-exploitability}.

We next show that the optimizer's coarse correlated-equilibrium benchmark in the constructed game is zero. Since the learners' utilities do not depend on the optimizer's action, the induced game among the learners is always the original bimatrix game $G$.
% Define
% \[
%     S_{\CE}(\widehat G)
%     :=
%     \operatorname{conv}
%     \left\{
%         y\otimes r:
%         y\in\Delta(A_0),\
%         r\in\CE(G)
%     \right\},
% \]
% Note, for all games, $S_\NE\subseteq S_\CE$, therefore any algorithm which approaches $S_\NE$ also approaches $S_\CE$.

% Next, define the optimizer's correlated-equilibrium response benchmark by
% \[
%     V_{\CE}
%     :=
%     \max_{\mu\in S_{\CE}(\widehat G)}
%     \left\langle
%         \widehat u_0,
%         \mu
%     \right\rangle.
% \]

\begin{lemma}
\label{lem:zero-ce-benchmark}
For the constructed game $\widehat G$,
\[
    V_{\CCE}=0.
\]
\end{lemma}

\begin{proof}
Fix any coarse correlated equilibrium $r\in\CCE(G)$. For every $i\in A_1$,
\begin{align*}
    \E_{(a_1,a_2)\sim r}
    \left[
        \widehat u_0(R_i,a_1,a_2)
    \right]
    &=
    \E_r
    \left[
        A_{i,a_2}-A_{a_1,a_2}
    \right]\le0.
\end{align*}
Indeed, this is the gain from the unconditional deviation in which player $1$
always switches to pure action $i$, and every coarse correlated equilibrium satisfies
all such deviation inequalities.

Likewise, for every $j\in A_2$,
\[
    \E_{(a_1,a_2)\sim r}
    \left[
        \widehat u_0(C_j,a_1,a_2)
    \right]
    =
    \E_r
    \left[
        B_{a_1,j}-B_{a_1,a_2}
    \right]
    \le0.
\]
Therefore every optimizer mixed strategy
$y\in\Delta(A_0)$ satisfies
\(
    \left\langle
        \widehat u_0,
        y\otimes r
    \right\rangle
    \le0.
\)
By linearity, the same inequality holds throughout
$S_{\CCE}(\widehat G)$, and hence
\(
    V_{\CCE}\le0.
\)

Furthermore, the optimizer can always play its null action, which gives
identically zero payoff. Hence
\(
    V_{\CCE}\ge0.
\)
Combining the two inequalities gives
\(
    V_{\CCE}=0.
\)
\end{proof}

We can now formally state our hardness result.

\begin{theorem}[Hardness of universal non-manipulability]
\label{thm:universal-nonmanip-hardness}
Suppose there exist universal algorithms for two independent learners with the following property: for every finite three-player game,
every adaptive optimizer%\footnote{Specifically, the hardness results hold even when the learners know the optimizer payoff before the game starts.}
, and every
$\varepsilon>0$, after
\(
    T=T(N,\varepsilon)
\)
rounds the optimizer's expected average payoff is at most
\(
    V_{\CCE}+\varepsilon,
\)
where $N$ denotes the encoding size of the three-player game.

Then, given an arbitrary bimatrix game $G$, there is a randomized algorithm
using these learner algorithms that outputs, with probability at least
$1/2$, a $2\varepsilon$-approximate Nash equilibrium of $G$ using
$T(N,\varepsilon)$ rounds of the learner algorithms and polynomial additional
computation per round.

More generally, for every $\delta\in(0,1)$, the success probability can be
increased to at least $1-\delta$ using
\(
    O(\log(1/\delta))
\)
independent repetitions.

Consequently, if the learner algorithms run in polynomial time per round and
$T(N,\varepsilon)$ is polynomial in $N$ and $1/\varepsilon$, then
inverse-polynomially accurate approximate Nash equilibria of bimatrix games
can be computed in randomized polynomial time. Such algorithms don't exist unless $\mathsf{PPAD} \subseteq \mathsf{FBPP}$.
\end{theorem}

The proof of this theorem follows straightforwards from Lemmas \ref{lem:optimizer-exploitability} and \ref{lem:zero-ce-benchmark}, we defer further details to Appendix \ref{app:theorem4.3proof}

\section{Characterization of Jointly Approachable Convex Sets}
\label{sec:characterization-of-approachable-sets}

The tools we developed in previous sections allow us to reason about properties of convex sets that go beyond non-manipulability. More generally, they provide an extension of Blackwell approachability to games with more than two players. 

In classical approachability theory, the goal of a learning algorithm is to ensure that the average of some vector-valued statistic of play approaches a desired target set. Formally, let $g\to\mathbb{R}^d$ map each joint action profile to a vector of statistics. The learner's goal is to ensure that the average statistic $\frac{1}{T}\sum_{t=1}^T g(a^{(t)})$ approaches a target set $S\subseteq\mathbb{R}^d$, regardless of the opponent's behavior. In the preceding sections, we showed that a particular target set of empirical action distributions (the joint menu) is approachable by multiple independent learners. Here, we generalize this question to arbitrary vector-valued statistics of play and characterize which convex target sets are jointly approachable without shared randomness.

We begin by providing the standard definition of approachability, and then the natural multi-learner extension:  
\begin{definition}
In a two-player repeated game, a compact set $S\subseteq\mathbb{R}^d$
is \emph{approachable} by the learner if there exists a learning
algorithm $L$ such that, against every optimizer policy $\pi$,
\[
\lim_{T\to\infty}
\mathbb{E}\left[\operatorname{dist}(\bar g_T,S)\right]=0.
\]
\end{definition}

\begin{definition}
In a repeated game with multiple learners, a compact set
$S\subseteq\mathbb{R}^d$ is \emph{jointly approachable} by the
learners if there exists a profile of independent learning algorithms
$\mathcal{A}_{\mathcal{L}}$ such that, against every optimizer policy
$\pi$,
\[
\lim_{T\to\infty}
\mathbb{E}\left[\operatorname{dist}(\bar g_T,S)\right]=0.
\]
\end{definition}

In classical approachability theory we know that a closed, convex set is approachable if and only if it satisfies two conditions: halfspace satisfiability and response satisfiability. These conditions are equivalent for closed, convex sets (\cite{Blackwell1956,AbernethyBartlettHazan2011}). %\jingyi{ (Need citations for this and this section in general.)} 
When extending this to n-player games without shared randomness, halfspace approachability remains the characterizing condition, while response-based approachability does not. In this section we illuminate this distinction. 

\paragraph{Two-player Blackwell approachability conditions.}
We begin by stating the two conditions of Blackwell approachability for two-player games, both of which equivalently characterize when a closed, convex set is approachable. 

Let player $0$ be the optimizer and player $1$ be the learner. We extend
$g$ multilinearly to mixed strategies, so that $g(y,x)$ denotes the
expected vector payoff when the optimizer plays $y$ and the learner plays $x$.

\begin{definition}[Halfspace satisfiability]
A compact set $S\subseteq\mathbb{R}^d$ is
\emph{halfspace satisfiable} if
\[
\forall U\in\mathbb{R}^d,\;
\exists x\in\Delta(A_1),\;
\forall y\in\Delta(A_0):
\quad
\langle U,g(y,x)\rangle
\leq \max_{s\in S}\langle U,s\rangle.
\]
\end{definition}

\begin{definition}[Response satisfiability]
A compact set $S\subseteq\mathbb{R}^d$ is
\emph{response satisfiable} if
\[
\forall y\in\Delta(A_0),\;
\exists x\in\Delta(A_1):
\quad g(y,x)\in S.
\]
\end{definition}

Halfspace satisfiability requires that, for every direction, the learner
can choose a strategy that guarantees the expected payoff lies in a
halfspace containing $S$, regardless of the optimizer's action. Response satisfiability instead requires that, for every mixed strategy
of the optimizer, the learner has a response whose expected payoff lies
in $S$. For convex and compact sets, these conditions are equivalent by
minimax and characterize approachability.

\paragraph{Multi-learner Blackwell approachability conditions.}
We now describe the natural extensions of these two conditions to the setting with multiple
independent learners. Since the learners do not share randomness,
their joint mixed strategy must be a product distribution
\[
q=\bigotimes_{i\in\learners}x_i,
\qquad x_i\in\Delta(A_i).
\]
For an optimizer mixed strategy $y\in\Delta(A_0)$, let
$g(y,q)$ denote the expected vector payoff under the
joint distribution $y\otimes q$.

\begin{definition}[Joint halfspace satisfiability]
A compact set $S\subseteq\mathbb{R}^d$ is
\emph{jointly halfspace satisfiable} if
\[
\forall U\in\mathbb{R}^d,\;
\exists q\in\bigotimes_{i\in\learners}\Delta(A_i),\;
\forall y\in\Delta(A_0):\quad
\langle U,g(y,q)\rangle
\leq \max_{s\in S}\langle U,s\rangle.
\]
\end{definition}

\begin{definition}[Joint response satisfiability]
A compact set $S\subseteq\mathbb{R}^d$ is
\emph{jointly response satisfiable} if
\[
\forall y\in\Delta(A_0),\;
\exists q\in\bigotimes_{i\in\learners}\Delta(A_i):\quad
 g(y,q)\in S.
\]
\end{definition}

Unlike in the two-player setting, the set of feasible joint
strategies $q$ is generally nonconvex. Thus the
classical minimax argument does not establish equivalence of
these two conditions. As we show below, for convex and compact
$S$, the joint halfspace satisfiability still characterizes joint
approachability, while joint response satisfiability does not. 

\begin{theorem}\label{thm:joint-approachability-characterization}
A convex and compact set $S\subseteq\mathbb{R}^d$ is jointly approachable by a set of independent learners $\learners$ if and only if it is jointly halfspace satisfiable.
\end{theorem}

The proof follows the same geometric argument used in our construction for non-manipulability. If the halfspace condition fails, the optimizer can choose an action that keeps the average payoff separated from $S$ in some direction. Conversely, if the condition holds, the learners can select a product distribution in each round that prevents the accumulated discrepancy from increasing in that direction. Although the learners sample their actions independently, they can coordinate these choices using the public history of optimizer actions. The construction obtains an $O(T^{-1/2})$ approachability rate for any fixed finite game. We give the algorithm and proof in Appendix~\ref{app:joint-approachability-proof}.

Finally, we also show that joint response satisfiability is neither necessary nor sufficient for joint approachability.

\begin{lemma}
    \label{lem:RSdneApproachability}
    There exist convex and compact sets $S\subseteq\mathbb{R}^d$ that are jointly approachable but not jointly response satisfiable. There also exist convex and compact sets $S\subseteq\mathbb{R}^d$ which are jointly response satisfiable but not jointly approachable. 
\end{lemma}

We defer the proof of this lemma to Appendix$~\ref{app:JRSdneJA}$.

Taken together, these results highlight a fundamental distinction between approachability by a single learner and by multiple independent learners. While the classical equivalence between halfspace and response satisfiability no longer holds without shared randomness, halfspace satisfiability continues to characterize exactly which convex sets are approachable. %In particular, even when the learners cannot produce a desired outcome through a single product strategy, they may still achieve it in the long run by varying their independent strategies over time.

\newpage

\bibliographystyle{plainnat}
\bibliography{bibliography}

\section*{Acknowledgments}

This material is based upon work supported by the National Science
Foundation CISE Graduate Fellowships under Grant No.~2313998.
Any opinions, findings, and conclusions or recommendations expressed
in this material are those of the author(s) and do not necessarily
reflect the views of the National Science Foundation.
\newpage
\appendix
\section{Appendix}

\subsection{AI Disclosure}

With respect to the central ideas of the paper, the authors decided that extending the notion of non-manipulability beyond two-player games was an interesting research direction. Based on authors' prior experience, we began by using the menu framework and approachability theory. We defined the menu that we wanted to approach to achieve both of our desired guarantees, but at first were unsure what algorithm would be able to approach it. At this point, the authors began discussing the problem with ChatGPT 5.6 extensively and provided the referenced related work along with a description of the algorithm which we wanted to produce. ChatGPT 5.6 produced an initial version of the algorithm which contained the key insights needed. The authors went on to make the initially produced algorithm more intuitive and digestible.

After developing our algorithm, the authors wondered if a significantly more efficient algorithm was possible given that our algorithm requires the computing of a Nash equilibrium at every timestep. This began an extensive dual track process of attempting to either develop more computationally efficient algorithms or produce a matching hardness result. Eventually, we produced this hardness result during an in-person working session which incorporated the frequent use of ChatGPT 5.6.

At this point, the authors noticed that the techniques we had developed seem to be more general than our particular setting. This led to us drafting a note that our algorithm relies upon a natural extension of Blackwell's halfspace satisfiability condition. Being familiar with approachability theory, this led the authors to question if there existed an equivalent extension of response satisfiability. This was later resolved as well. Again, ChatGPT 5.6 was used as a tool throughout this leg of the research process. The authors used the tool iteratively, it required several rounds of prompting, refining our questions, and providing specific context to arrive at our results. Some particularly simple proofs were one-shot by humans.

With regards to writing, the general pipeline we used was to construct a detailed outline, provide it to ChatGPT 6 to fill in details, and then extensively refine the results. The initial AI-generated writing was often overly verbose, used meaningless ornamental language, and lacked helpful intuition. A significant part of the refinement process was removing vague or overly verbose language, and adding intuition. Several AI-generated proofs to lemmas were ten times longer than they needed to be and required extensive trimming. ChatGPT 6 was also used for final technical spot checks for minor bugs. For this as well, however, the list of issues it identified was significantly longer and more pedantic than was required. We have sufficiently addressed all issues of importance.

\subsection{Example of no swap-regret manipulable algorithms}
\label{app:counterexample}

In this section, we give a short example which shows that if all learners play an algorithm that guarantees they personally have no-swap regret, they can nevertheless be jointly manipulable. We do so by the following lemma.

\begin{lemma}%[No-swap regret learners can be jointly manipulable]
\label{lem:BMisManipulable}
The Blum-Mansour algorithm, which guarantees no-swap regret, is jointly manipulable.
\end{lemma}

\begin{proof}
All players have action set $\{0,1\}$, and their utilities are
\[
    u_i(c,a_1,a_2)=\mathbf{1}\{a_i=c\},
    \qquad i\in\{1,2\},
    \qquad
    u_0(c,a_1,a_2)=\mathbf{1}\{a_1\ne a_2\}.
\]
Thus, both learners want to match the optimizer, whereas the optimizer
wants the learners to disagree. Observing the optimizer's action
reveals each learner's entire payoff vector.

We specify the exponential-weights instantiation of Blum--Mansour.
Each learner initializes $W_1(a,b)=1$ for $a,b\in\{0,1\}$.
In round $t$, form
\[
    Q_t(a,b)
    =
    \frac{W_t(a,b)}{\sum_{r=0}^1 W_t(a,r)}
\]
and play the unique stationary distribution
$x_t=x_tQ_t$.
After observing $c^{(t)}$, update
\[
    W_{t+1}(a,b)
    =
    W_t(a,b)
    \exp\!\left(
        \eta x_t(a)\mathbf{1}\{b=c^{(t)}\}
    \right).
\]
The two learners therefore use the same mixed action
$x_t=(1-p_t,p_t)$, but sample their actions independently.

For each row $a$, the exponential-weights regret bound on any
prefix of length $r\le T$ is
\[
    \frac{\log 2}{\eta}
    +
    \frac{\eta}{8}\sum_{t=1}^r x_t(a)^2,
\]
because its round-$t$ payoff vector has range $x_t(a)$.
Summing the two row bounds and using $x_tQ_t=x_t$ gives
\begin{equation}
\label{eq:bm-example-prefix-regret}
    \max_{\phi:\{0,1\}\to\{0,1\}}
    \sum_{t=1}^r\sum_{a=0}^1
    x_t(a)
    \left(
        \mathbf{1}\{\phi(a)=c^{(t)}\}
        -
        \mathbf{1}\{a=c^{(t)}\}
    \right)
    \le
    \frac{2\log 2}{\eta}+\frac{\eta r}{8}
    \le 2\sqrt{T}.
\end{equation}
This bound holds for every optimizer-action sequence.

For each of the four swaps, its realized cumulative gain minus its
mixed-action cumulative gain is a martingale with second moment
at most $T$. By Cauchy--Schwarz, the expected maximum of these
four differences is at most $2\sqrt{T}$. Consequently,
\[
    \mathbb{E}[R_i^{\mathrm{sw}}(T)]
    \le 4\sqrt{T},
    \qquad i\in\{1,2\}.
\]
This guarantee holds against arbitrary admissible opponents.

Set
\[
    d_t:=2c^{(t)}-1,
    \qquad
    \alpha_t:=\eta\sum_{s<t}(1-p_s)d_s,
    \qquad
    \beta_t:=\eta\sum_{s<t}p_s d_s,
    \qquad
    Z_t:=\alpha_t+\beta_t
        =\eta\sum_{s<t}d_s.
\]
Writing $\sigma(z):=(1+e^{-z})^{-1}$, the weight updates give
\[
    Q_t(0,1)=\sigma(\alpha_t),
    \qquad
    Q_t(1,0)=\sigma(-\beta_t).
\]
Hence stationarity implies
\[
    p_t
    =
    \frac{\sigma(\alpha_t)}
         {\sigma(\alpha_t)+\sigma(-\beta_t)},
    \qquad
    p_t-\frac12
    =
    \frac{\sinh(Z_t/2)}
         {2\bigl(
             \cosh(Z_t/2)+e^{(\alpha_t-\beta_t)/2}
          \bigr)}.
\]
The mixed-action cumulative gains of the two one-way swaps
$0\mapsto1$ and $1\mapsto0$ are respectively
$\alpha_t/\eta$ and $-\beta_t/\eta$.
Applying \eqref{eq:bm-example-prefix-regret} to the prefix
$t-1$ therefore gives
\[
    \alpha_t\le 2,
    \qquad
    -\beta_t\le 2.
\]
In particular, $e^{(\alpha_t-\beta_t)/2}\le e^2$, so
\begin{equation}
\label{eq:bm-example-imbalance}
    \frac{\tanh(|Z_t|/2)}{2(1+e^2)}
    \le
    \left|p_t-\frac12\right|
    \le
    \frac12\tanh(|Z_t|/2)
    \le
    \frac{|Z_t|}{4}.
\end{equation}

Since the learners sample independently, the optimizer's conditional
expected payoff in round $t$ is
\[
    2p_t(1-p_t)
    =
    \frac12-2\left(p_t-\frac12\right)^2.
\]
Write $V_T(\pi)$ for the optimizer's expected average payoff
under policy $\pi$.

%\paragraph{A dynamic optimizer.}
Let the optimizer alternate deterministically:
\[
    c^{(1)},c^{(2)},c^{(3)},\ldots
    =
    0,1,0,1,\ldots.
\]
Every prefix has imbalance at most one, so $|Z_t|\le\eta$.
By \eqref{eq:bm-example-imbalance},
$|p_t-\tfrac12|\le\eta/4$. Thus
\begin{equation}
\label{eq:bm-example-dynamic}
    \sup_{\pi^T\in\Pi^T}V_T(\pi^T)
    \ge
    \frac12-\frac{\eta^2}{8}
    =
    \frac12-\frac{1}{8T}.
\end{equation}

%\paragraph{Every static optimizer.}
Fix any static distribution $y$, and write
$\theta:=y(1)$ and $\mu:=2\theta-1$.
Under $y^{\otimes T}$, the variables $d_t$ are independent,
take values in $\{-1,1\}$, and have mean $\mu$.
For
\[
    S_n:=\sum_{s=1}^n d_s,
    \qquad
    v_n:=\mathbb{E}[S_n^2],
\]
direct expansion gives
\[
    v_n=n+n(n-1)\mu^2\ge n
\]
and
\begin{align*}
    \mathbb{E}[S_n^4]
    &=
    3n^2-2n
    +2n(n-1)(3n-4)\mu^2 \\
    &\qquad
    +n(n-1)(n-2)(n-3)\mu^4
    \le 3v_n^2.
\end{align*}
Applying the Paley--Zygmund inequality to $S_n^2$ yields
\[
    \Pr\!\left(
        |S_n|\ge\sqrt{v_n/2}
    \right)
    \ge
    \frac{v_n^2}{4\mathbb{E}[S_n^4]}
    \ge \frac1{12}.
\]
Therefore, whenever $n=t-1\ge T/2$,
\[
    \Pr\!\left(|Z_t|\ge\frac12\right)
    =
    \Pr\!\left(\eta|S_n|\ge\frac12\right)
    \ge \frac1{12}.
\]
Define the universal constant
\[
    \kappa
    :=
    \frac{\tanh(1/4)}{2(1+e^2)}
    >0.
\]
By \eqref{eq:bm-example-imbalance}, every such round satisfies
\[
    \mathbb{E}\!\left[
        \left(p_t-\frac12\right)^2
    \right]
    \ge \frac{\kappa^2}{12}.
\]
There are $\lfloor T/2\rfloor\ge T/3$ such rounds for $T\ge2$.
Consequently,
\[
    V_T(y^{\otimes T})
    =
    \frac12
    -
    \frac2T\sum_{t=1}^T
    \mathbb{E}\!\left[
        \left(p_t-\frac12\right)^2
    \right]
    \le
    \frac12-\frac{\kappa^2}{18}.
\]
This bound is uniform over $\theta\in[0,1]$, including choices
depending on $T$.

Combining this bound with \eqref{eq:bm-example-dynamic} gives
\[
    \sup_{\pi^T\in\Pi^T}V_T(\pi^T)
    -
    \sup_{y\in\Delta(A_0)}V_T(y^{\otimes T})
    \ge
    \frac{\kappa^2}{18}-\frac1{8T}.
\]
For all sufficiently large $T$, the right-hand side is at least
$\kappa^2/36>0$, proving joint manipulability.
\end{proof}

\subsection{Extended Related Work}
\label{sec:ext-related-work}
\paragraph{Non-manipulability.}
\citet{BravermanMaoSchneiderWeinberg2018} initiated the study of manipulating no-regret learners by demonstrating that when a buyer uses a broad class of mean-based no-regret
algorithms, a strategic seller can manipulate the buyer's learning dynamics
and extract revenue approaching the available welfare.
\citet{CaiWeinbergWildenhainZhang2023} extended this result to any number of i.i.d.\ mean-based buyers.
\citet{DengSchneiderSivan2019} introduced the first formal notion of non-manipulability by showing that an
optimizer cannot
asymptotically exceed its Stackelberg value against a no-swap-regret
learner. \citet{MansourMohriSchneiderSivan2022} extend this framework to Bayesian and more general polytope games.
They introduce polytope swap regret and study the payoff guarantees that
a strategic optimizer can force and the ability of a learner to limit the
optimizer to those benchmarks.

More recent work has studied non-manipulability in auction environments, whose particular structure can be leveraged to enable efficient non-manipulability guarantees. \cite{KumarSchneiderSivan2024} first showed that online gradient descent in the quantile strategy space obtains low-regret and non-manipulability for learners. More recently, \cite{Zhao2026} and \cite{CaiLuoWeiZheng2026} extend this to show that the same guarantees can be obtained by any sublinear linearized regret algorithm operating in the quantile strategy space. In contrast to this work, we give algorithms and hardness results for the more general setting of $n$-player normal form games.

\paragraph{Menus.}

\citet{ArunachaleswaranCollinaSchneider2024} introduce the
\emph{asymptotic menu} of a learning algorithm: the set of long-run correlated
strategy profiles that an optimizer can induce against that algorithm. The
menu directly captures the algorithm's strategic vulnerability because it
describes the full set of outcomes available to an optimizer that understands
and responds to the learner's dynamics.

\citet{ArunachaleswaranCollinaMansourMohriSchneiderSivan2025} sharpen the
connection between regret and non-manipulability beyond normal-form games.
They introduce profile swap regret for polytope games and show that sublinear
profile swap regret is necessary and sufficient for non-manipulability against
a dynamic optimizer. They also give efficient algorithms for minimizing
profile swap regret, despite the computational difficulty of evaluating the
quantity on a fixed transcript. The survey of
\citet{ArunachaleswaranCollinaMansourMohriSivanSchneider2026} synthesizes
this literature through the menu framework.% and explains how swap regret functions as an optimal commitment rule against strategic opponents.

\paragraph{Approachability.}
The technical core of this paper leverages and extends Blackwell approachability techniques. Blackwell's approachability theorem gives a geometric characterization of when
a player in a repeated vector-payoff game can force the average payoff toward
a convex target set \citep{Blackwell1956}. Regret minimization can be encoded
directly as such a problem. In a loss-minimization formulation, for a finite
family of deviations \(\Phi\), define the instantaneous deviation-gain vector
by
\[
    g^{(t)}(\phi)
    =
    \ell^{(t)}(p^{(t)})-\ell^{(t)}\bigl(\phi(p^{(t)})\bigr),
    \qquad
    \phi\in\Phi.
\]
Approaching the negative orthant
\(\mathbb R_-^{|\Phi|}\) guarantees that every coordinate of the average
gain vector is asymptotically nonpositive and therefore that
\[
    \max_{\phi\in\Phi}
    \sum_{t=1}^T
    \left[
        \ell^{(t)}(p^{(t)})-\ell^{(t)}\bigl(\phi(p^{(t)})\bigr)
    \right]
    =
    o(T).
\]
External, swap, and more general \(\Phi\)-regret objectives arise
from different choices of the deviation family and vector-payoff
representation.

\citet{GordonGreenwaldMarks2008} develop a broad
\(\Phi\)-regret framework over convex action sets. \citet{AbernethyBartlettHazan2011} formalize the close algorithmic
relationship between Blackwell approachability and no-regret online linear
optimization.

Classical Blackwell algorithms choose the approachability direction by
projecting the current average payoff onto the target set.
\citet{BernsteinShimkin2015} instead develop response-based approachability algorithms
from Blackwell's dual condition.

\citet{DannMansourMohriSchneiderSivan2025} refine the
approachability--regret connection by showing that generic reductions need not
preserve optimal rates and by identifying improper \(\Phi\)-regret as a
rate-tight intermediary. In general convex games,
\citet{DaskalakisFarinaFishelsonPipisSchneider2025} give the first
polynomial-time methods for minimizing linear swap regret and computing linear
correlated equilibria.
\citet{AnagnostidesFarinaFishelsonLuoSchneider2026} combine
response-based approachability with geometric preconditioning to obtain an
\(O(d\sqrt T)\) linear-swap-regret algorithm over general convex sets,
simultaneously controlling profile swap regret, and prove a matching
information-theoretic lower bound over centrally symmetric domains.

\subsection{Handling Learner Defection}
\label{app:learner-defection}

In this section, we show how to modify our algorithm to guarantee that each learner has no-swap regret even if other learners defect from the agreed upon algorithm. The key idea of our modification is that we will have each player simulate the behavior of the other players to measure what their expected utility should be and if they find that their realized utility is significantly less than what they expect then they switch to running the \cite{BlumMansour2007} algorithm with bandit feedback.

%The key idea of our modification is that we will augment the coordinated protocol with independent statistical tests and an importance-weighted implementation of the Blum--Mansour algorithm. Each learner retains no swap regret even if the other learners behave arbitrarily, while an adaptive optimizer cannot cause a switch under universal learner compliance except with a prescribed small probability.

%The detection guarantee concerns statistically significant departures that affect a learner's swap gains. A change of behavior that induces exactly the same feedback distribution cannot be identified from these observations. Likewise, the absence of false alarms is a high-probability guarantee: finite observations with overlapping distributions do not permit universal detection with zero probability of error.

\subsubsection*{1. Protocol Description}

Recall that, after each round, learner $i$ observes the optimizer's
realized action $c^{(t)}$ and her own realized utility
\[
    r_i^{(t)}
    :=u_i(c^{(t)},a_i^{(t)},a_{-i}^{(t)}).
\]
She need not observe the other learners' actions, realized utilities, or
private randomness. The learner utility functions remain known, as in
Section~3. 

The optimizer may adapt to the entire past and to the learners' current
mixed actions, as in Section~3. All current actions of other players must
be chosen without observing the fresh sampled action of the learner whose
regret is being protected. Adhering learners sample their actions
independently using private randomness.

Recall
$\Phi_i:=\{\phi:A_i\to A_i\}$. Fix a horizon $T\ge2$ and an error parameter
$\alpha\in(0,1)$. Set
\[
    \gamma:=T^{-1/4},
    \qquad K_\Phi:=\sum_{j\in\mathcal L}|\Phi_j|,
    \qquad
    \Lambda_T:=\log\frac{6TK_\Phi}{\alpha},
\]
and define
\[
    M_i:=\frac{|A_i|}{\gamma}+2,
    \qquad h_i(s):=M_i\sqrt{2s\Lambda_T}.
\]
For asymptotic guarantees we take $\alpha=(T+1)^{-2}$.

Let $\mathcal A_{\mathcal L}^{\mathrm{base}}$ be the coordinated protocol.
Every learner independently simulates this protocol on the observed
optimizer-action history, obtaining prescribed distributions
$\widehat x_j^{(t)}$ and
\(
    \widehat q^{(t)}:=\bigotimes_{j\in\mathcal L}\widehat x_j^{(t)}.
\)
This virtual simulation continues to use the original prescribed
distributions in its state updates. %In particular, the virtual update in Algorithm~1 uses $\widehat q^{(t)}$, even though actual play will include exploration.

We use the following uniform prefix guarantee of the base protocol:
for every optimizer-action sequence and every $s\le T$,
\begin{equation}
    \max_{\phi\in\Phi_i}
    \sum_{t=1}^{s}
       G_{i,\phi}(e_{c^{(t)}}\otimes\widehat q^{(t)})
    \le B_i(T),
    \qquad B_i(T)=o(T).
    \label{eq:defection-base-prefix}
\end{equation}
For Algorithm~1, Lemma~\ref{lem:product-play-approachability} together with the swap-gain inequalities and norm bound give
\[
    B_i(T)=K_i\sqrt{2T},
    \qquad K_i:=\max_{\phi\in\Phi_i}\|G_{i,\phi}\|_2.
\]
%The application to the completed protocol is discussed in part~2.

%\paragraph{Monitoring mode.}

Our modified algorithm requires that each learner slightly mixed the recommended distribution from the original algorithm with the uniform distribution. The purpose of this is to ensure that every action has positive probability such that we can construct useful estimates for the payoffs of all actions, which will allow us to measure whether other learners are defecting. The mixed distribution which we have each learner play is as follows
\[
    \overline p_j^{(t)}
      :=(1-\gamma)\widehat x_j^{(t)}
        +\gamma\,\operatorname{Unif}(A_j),
    \qquad
    \overline q^{(t)}
      :=\bigotimes_{j\in\mathcal L}\overline p_j^{(t)}.
\]
An immediate observation is that playing these distributions ensures that every action of every player has probability of being selected at least $\gamma/|A_i|$.

After observing $c^{(t)}$ and $r_i^{(t)}$, learner $i$ computes the nominal
payoff vector
\[
    \nu_i^{(t)}(b)
      :=u_i(c^{(t)},b,\overline p_{-i}^{(t)}),
      \qquad b\in A_i,
\]
where $\overline p_{-i}^{(t)}$ denotes the product of the other learners'
prescribed monitoring distributions and $c^{(t)}$ is the optimizer's actual realized action.

The learner also forms estimates of the utility of each action as follows
\[
    \widehat v_i^{(t)}(b)
      :=\frac{r_i^{(t)}\mathbf 1\{a_i^{(t)}=b\}}
              {\overline p_i^{(t)}(b)}.
\]
Here, we weight each observation by the inverse of the probability of observing it so that $\hat{v}_i^{(t)}(b)$ is an unbiased estimator of $ \nu_i^{(t)}(b)$ when all learners are complying.

For each $\phi\in\Phi_i$, define
\begin{align*}
    \widehat g_{i,\phi}^{(t)}
      &:=\sum_{a\in A_i}\overline p_i^{(t)}(a)
           [\widehat v_i^{(t)}(\phi(a))-\widehat v_i^{(t)}(a)],\\
    \overline g_{i,\phi}^{(t)}
      &:=\sum_{a\in A_i}\overline p_i^{(t)}(a)
           [\nu_i^{(t)}(\phi(a))-\nu_i^{(t)}(a)]
       =G_{i,\phi}(e_{c^{(t)}}\otimes\overline q^{(t)}).
\end{align*}
The first quantity estimates the actual swap gain, and the second is the
swap gain predicted under compliance by the other learners.

Learner $i$ maintains
\[
    D_{i,\phi}(s)
       :=\sum_{t=1}^{s}
            [\widehat g_{i,\phi}^{(t)}-\overline g_{i,\phi}^{(t)}]
\]
At the first round $\tau_i$ for which
\begin{equation*}
    \max_{\phi\in\Phi_i}|D_{i,\phi}(\tau_i)|>h_i(\tau_i),
    \label{eq:defection-alarm}
\end{equation*}
she flags that a learner has defected and permanently switches to running the $SR\_MAB$ algorithm of \cite{BlumMansour2007}.

This protocol can be performed efficiently because it does not require that we enumerate $D_{i,\phi}$ for every $\phi$. Instead, we can maintain the $|A_i|\times |A_i|$ matrix
\[
    E_i^{(s)}(a,b)
       :=\sum_{t=1}^{s}\overline p_i^{(t)}(a)
          [\widehat v_i^{(t)}(b)-\widehat v_i^{(t)}(a)
           -\nu_i^{(t)}(b)+\nu_i^{(t)}(a)].
\]
Then $D_{i,\phi}(s)=\sum_a E_i^{(s)}(a,\phi(a))$, and
\[
    \max_{\phi}|D_{i,\phi}(s)|
      =\max\left\{
          \sum_a\max_b E_i^{(s)}(a,b),
          -\sum_a\min_b E_i^{(s)}(a,b)
        \right\}.
\]
This uses $O(|A_i|^2)$ stored values and arithmetic per
round, apart from the computation of the virtual protocol.

%\paragraph{Blum--Mansour fallback.}
%The usual full-information Blum--Mansour algorithm cannot be supplied with unobserved counterfactual payoffs. We use its reduction on bounded importance-weighted estimates instead.

%On entering fallback, learner $i$ starts a fresh Blum--Mansour instance for the remaining horizon. If the instance outputs $w_i^{(t)}$, she actually samples from
%\[
%    p_i^{(t)}
%       :=(1-\gamma)w_i^{(t)}
%         +\gamma\,\operatorname{Unif}(A_i).
%\]
%After observing her utility, she forms
%\[
%    \widetilde\ell_i^{(t)}(b)
%       :=\frac{(1-r_i^{(t)})\mathbf1\{a_i^{(t)}=b\}}
%                    {2p_i^{(t)}(b)},
%    \qquad
%    z_i^{(t)}:=\frac{\gamma}{|A_i|}\widetilde\ell_i^{(t)}.
%\]
%Since $p_i^{(t)}(b)\ge\gamma/|A_i|$, the supplied vector $z_i^{(t)}$ lies in $[0,1]^{|A_i|}$. Feed $z_i^{(t)}$ to the Blum--Mansour instance. More explicitly, if its external-regret copies produce the rows of $Q_i^{(t)}$, choose $w_i^{(t)}=w_i^{(t)}Q_i^{(t)}$ and give copy $b$ the loss vector $w_i^{(t)}(b)z_i^{(t)}$. The weights in this internal update are the stationary probabilities $w_i^{(t)}(b)$. This fallback uses only the learner's own action and realized utility.

\subsubsection*{2. Detection of Learner Defection}

For analysis, let $\mathcal F_t$ contain all past play, the current mixed
actions, and $c^{(t)}$, before the learners' fresh actions are sampled.
For any learner $i$ still monitoring, define the actual conditional payoff
vector
\[
    v_i^{(t)}(b)
       :=\mathbb E[u_i(c^{(t)},b,a_{-i}^{(t)})\mid\mathcal F_t]
\]
and the actual conditional swap gain
\[
    \mu_{i,\phi}^{(t)}
       :=\sum_a\overline p_i^{(t)}(a)
             [v_i^{(t)}(\phi(a))-v_i^{(t)}(a)].
\]
Define the cumulative
swap-relevant departure from prescribed play by
\[
    \Delta_{i,\phi}(s)
       :=\sum_{t=1}^{s}
            [\mu_{i,\phi}^{(t)}-\overline g_{i,\phi}^{(t)}].
\]

\begin{lemma}%[Detection of statistically significant defection]
\label{lem:defection-detection}
With probability at least $1-\alpha/3$, simultaneously for all adhering
learners, all swaps, and all monitoring prefixes,
\begin{equation*}
    |D_{i,\phi}(s)-\Delta_{i,\phi}(s)|\le h_i(s)=O(T^{1/4}\sqrt{s\log T}).
    \label{eq:defection-drift-concentration}
\end{equation*}
Consequently, if a monitoring prefix satisfies
$|\Delta_{i,\phi}(s)|>2h_i(s)$ for some swap $\phi$, learner $i$
swaps to the $SR\_MAB$ algorithm in the following round. Hence, the modified algorithm guarantees every learner sublinear swap-regret.
\end{lemma}

\begin{proof}
Independence of learner $i$'s action from the other players' 
actions implies
\[
    \mathbb E[\widehat v_i^{(t)}(b)\mid\mathcal F_t]
       =v_i^{(t)}(b),
    \qquad
    \mathbb E[\widehat g_{i,\phi}^{(t)}\mid\mathcal F_t]
       =\mu_{i,\phi}^{(t)}.
\]
%These identities hold even when the other learners deviate and the optimizer adapts to the current distributions.

Writing $b=a_i^{(t)}$, the unbiased estimator of swap gain can be rewritten as
\[
    \widehat g_{i,\phi}^{(t)}
       =r_i^{(t)}
          \left(
           \frac{\sum_{a:\phi(a)=b}\overline p_i^{(t)}(a)}
                {\overline p_i^{(t)}(b)}-1
          \right).
\]

To apply the Azuma-Hoeffding bound, we define the centered increment as $X_t=g_{i,\phi}^{(t)}-\mu_{i,\phi}^{(t)}$.

Since $\overline{p}^{(t)}_i=(1-\gamma)\hat{x}_i^{(t)}+\gamma\text{Unif}(A_i)$, we have $\overline{p}_i^{(t)}(b)\geq\gamma/|A_i|$. Therefore $|\widehat g_{i,\phi}^{(t)}|\le |A_i|/\gamma$.

Since $|\mu_{i,\phi}^{(t)}|\le2$, the centered increments have absolute
value $|X_t|\leq |A_i|/\gamma+2= M_i$.

Then for any $s$ round monitoring prefix and for each $i,\phi$, Azuma--Hoeffding gives
\[
    \Pr\left(
       \left|\sum_{t=1}^{s}
              (\widehat g_{i,\phi}^{(t)}-\mu_{i,\phi}^{(t)})
       \right|>h_i(s)
       \right)
       \le 2e^{-\Lambda_T}
       =\frac{\alpha}{3TK_\Phi}
\]
Finally, union bounding over the at most
$TK_\Phi$ learner--swap--time combinations gives
$$|D_{i,\phi}(s)-\Delta_{i,\phi}(s)|\le h_i(s)=O(T^{1/4}\sqrt{s\log T}).$$

\end{proof}

\subsubsection*{3. No Switching Without Learner Defection}

Lastly in this section, we show that the modified algorithm maintains the properties of non-manipulability

\begin{lemma}%[Protection against false alarms caused by the optimizer]
\label{lem:defection-no-false-alarm}
For every optimizer policy, if all learners follow the modified algorithm, then
\[
    \Pr(\text{some learner rejects the compliance test by time }T)
       \le\alpha/3=O(T^{-2}).
\]
\end{lemma}

\begin{proof}
Suppose every learner always uses
$\overline p_j^{(t)}$ and never enters $SR\_MAB$. This reference
still includes the prescribed exploration of all actions available to the learner.

Condition on $\mathcal F_t$, which includes the optimizer's realized
action $c^{(t)}$. Independent sampling by the learners gives
\[
    \mathbb E[\widehat v_i^{(t)}(b)\mid\mathcal F_t]
       =u_i(c^{(t)},b,\overline p_{-i}^{(t)})
       =\nu_i^{(t)}(b).
\]
Consequently,
\[
    \mathbb E[
       \widehat g_{i,\phi}^{(t)}
          -\overline g_{i,\phi}^{(t)}
       \mid\mathcal F_t]=0.
\]
Every $D_{i,\phi}$ is therefore a martingale with increments bounded
by $M_i$. Azuma--Hoeffding gives
\[
    \Pr(|D_{i,\phi}(s)|>h_i(s))
       \le\frac{\alpha}{3TK_\Phi}.
\]
A union bound over learners, swaps, and times shows that any reference
test rejects with probability at most $\alpha/3$.

Couple the actual execution with this reference execution using the
same optimizer policy and the same random samples whenever their
histories agree. The executions coincide up to and including the
first rejected test, so their first-rejection times coincide.
Consequently, the probability of any rejection in the actual
execution is also at most $\alpha/3$.
\end{proof}

Let $A_L^{\mathrm{base}}$ be the menu-completed protocol,
write $A_L^{\mathrm{mod}}$ for its monitored modification, and fix
$\|u_0\|_\infty\le 1$ and
$V:=\max_{s\in S_{\NE}}\langle u_0,s\rangle$.
In the reference execution, exploration changes each conditional
expected payoff by at most $2|L|\gamma$, since
\[
    \left\|
        \bigotimes_{j\in L}\overline p_j^{(t)}
        -\widehat q^{(t)}
    \right\|_1
    \le
    \sum_{j\in L}
        \|\overline p_j^{(t)}-\widehat x_j^{(t)}\|_1
    \le 2|L|\gamma.
\]
Moreover, the preceding coupling shows that, for each optimizer
policy, the actual and reference expected average payoffs differ
by at most $2\alpha/3$.
Applying the base protocol's pathwise containment guarantee to the
optimizer-action sequence in the reference execution therefore gives
\[
    \sup_{\pi^T\in\Pi_T}
    \langle u_0,\phi_T(\pi^T,A_L^{\mathrm{mod}})\rangle
    \le V+o(1)+2|L|\gamma+\frac{2\alpha}{3}.
\]
Conversely, for a static optimizer, couple the base and reference
executions using the same i.i.d.\ optimizer actions; their virtual
base trajectories then agree, so static attainability for the base
protocol gives
\[
    \sup_{y\in\Delta(A_0)}
    \langle u_0,\phi_T(y^{\otimes T},A_L^{\mathrm{mod}})\rangle
    \ge V-o(1)-2|L|\gamma-\frac{2\alpha}{3}.
\]
Subtracting these inequalities bounds the dynamic-versus-static
payoff gap by
\[
    o(1)+4|L|\gamma+\frac{4\alpha}{3}=o(1),
\]
which proves joint non-manipulability.

\subsection{Proof of Lemma~\ref{lem:menu-criterion}}
\label{app:menu-criterion-proof}

\begin{proof}
We begin by showing the result for individual swap regret. Fix learner $i$. Let $\menu_T$ be the horizon-$T$ menu, and let
$\delta_T$ denote its Hausdorff distance from $\menu$.
By the definition of $\menu_T$, every optimizer-action sequence
induces a mixed-action empirical CSP $\bar k_T\in\menu_T$.
Every point of $\menu$ satisfies learner $i$'s swap inequalities,
and each swap-gain functional is linear and continuous. Hence,
uniformly over optimizer-action sequences,
\[
 \max_{\phi_i\in\Phi_i} G_{i,\phi_i}(\bar k_T)
 \le C_G\delta_T=o(1),
\]
where $C_G$ depends only on the fixed finite game.
Conditional on the past, the learners' current mixed strategies,
and the optimizer's realized action, the learners independently
sample their actions. For any fixed swap $\phi_i$, the difference
between its realized cumulative gain and its mixed-action
cumulative gain is therefore a bounded-increment martingale.
Since $\Phi_i$ is finite, standard martingale concentration gives
\[
 \sup_{\pi^T}\mathbb E[R_i^{\sw}(T)]
 \le C_G T\delta_T + O_G(\sqrt T)=o(T).
\]

Next we turn to joint non-manipulability. Fix $u_0:A\to[-1,1]$ and let $s^*$ be an extreme point of the
compact convex set $\menu$ maximizing $\langle u_0,s\rangle$.
Because the horizon-$T$ menus converge to $\menu$, every dynamic
optimizer policy has expected payoff at most
$\langle u_0,s^*\rangle+o(1)$.
By the second condition, some static optimizer strategy has expected
CSP within $o(1)$ of $s^*$, and hence obtains payoff at least
$\langle u_0,s^*\rangle-o(1)$.
Thus dynamic optimization improves the optimizer's expected average
payoff over its best static strategy by at most $o(1)$.
\end{proof}

\subsection{Menu Completion}\label{app:menu-completion}
% Proposed replacement for Appendix A.4.
% Retains a rational grid and approximate-Nash certification; no precomputed
% net of exact equilibria is assumed. Uses the paper's existing macros.

The purpose of this construction is to make every Nash-response outcome
attainable by a static optimizer while keeping every optimizer's average
outcome asymptotically inside $S_{\NE}$. The modified algorithm uses a finite list of approximate Nash responses, encodes each
list entry in a nearby optimizer distribution, and honors the decoded entry
until the optimizer's empirical play becomes inconsistent with it.
The implementing optimizer uses one distribution throughout the horizon.

Let $$E_{\NE}:=\{y\otimes q:q\in\NE(\gameL(y))\},\quad S_{\NE}:=\operatorname{conv}(E_{\NE}).$$
We use Algorithm~1's pathwise bound
$\operatorname{dist}_1(\bar k_r,S_{\NE})\le C_G/\sqrt r$
for every prefix of a fresh run, where $k^{(t)}=e_{c^{(t)}}\otimes q^{(t)}$.

\begin{lemma}%[Static attainability and menu completion]
\label{lem:menu-completion}Suppose $|A_0| >= 2$. Then Algorithm~1 admits a horizon-aware augmentation
$\alg'_{\learners}$ such that
\begin{align*}
 \sup_{\pi^T}\mathbb E\!\left[
   \operatorname{dist}_1(\bar k_T,S_{\NE})\right]&=o(1),\\
 \sup_{s\in E_{\NE}}\inf_{z\in\Delta(A_0)}
  \|\varphi_T(z^{\otimes T},\alg'_{\learners})-s\|_1&=o(1),\\
 \sup_{\pi^T}\mathbb E[R_i^{\sw}(T)]&=o(T)
 \qquad(i\in\learners).
\end{align*}
Consequently, the learners' asymptotic menu is $S_{\NE}$
and they are jointly non-manipulable.
\end{lemma}

\begin{proof}
%\noindent\textbf{1. A finite list and its labels.}
Set $\varepsilon=1/\lceil\log(T+2)\rceil$ and $m=\lceil T^{2/3}\rceil$.
Take a rational $\varepsilon$-net of polynomial size in $1/\varepsilon$
for the full strategy-profile space under
$\|y-y'\|_1+\sum_i\|x_i-x_i'\|_1$.
Retain precisely the grid profiles $(y_j,x_j)$ satisfying
\[
 u_i(y_j,b_i,x_{j,-i})-u_i(y_j,x_j)\le2\varepsilon
 \qquad\text{for every }i\text{ and }b_i\in A_i,
\]
and write $q_j=\bigotimes_i x_{j,i}$. For the fixed game, the retained list
has $N=(\log T)^{O_G(1)}$ entries. Rounding any exact Nash-response profile
by total error at most $\varepsilon$ changes each deviation gain by at most
$2\varepsilon$, so the list approximates every point of $E_{\NE}$ within
$\varepsilon$. Moreover,
\[
 d_T:=\max_j\operatorname{dist}_1(y_j\otimes q_j,E_{\NE})=o(1).
\]
Indeed, any contrary sequence has a convergent subsequence of strategy
profiles. The limit satisfies all Nash inequalities, so its joint outcome
belongs to $E_{\NE}$, a contradiction.

Fix an optimizer action $c_\star$ and let $M=\lceil4N/\varepsilon\rceil$.
Choose distinct $\theta_j\in\{0,1/M,\ldots,1\}$ with
$|\theta_j-y_j(c_\star)|\le\varepsilon/2$.
Greedy selection works because each admissible interval contains at least
$2N$ grid values. By moving mass to or from $c_\star$, choose a distribution
$z_j$ with
\[
 z_j(c_\star)=\theta_j,
 \qquad \|z_j-y_j\|_1\le\varepsilon.
\]
Thus one action's frequency identifies the list entry. All lists and
selection rules are fixed publicly; the checks above require only evaluating
pure deviations, not solving for exact equilibria.

We now define the modified algorithm as follows.

For the first $m$ rounds, the learners play fixed pure actions and estimate
the frequency $\widehat p$ of $c_\star$.
They decode $j$ if $|\widehat p-\theta_j|\le1/(3M)$; such a label is unique.
If decoding fails, they start a fresh run of Algorithm~1.
Otherwise, each learner independently plays her factor of $q_j$ if

\begin{equation*}
 \left\|\sum_{t=m+1}^{m+\ell}(e_{c^{(t)}}-z_j)\right\|_1\le m
 %\label{eq:mc-simple-monitor}
\end{equation*}
holds after every committed round. At the first failure, they start a fresh
run of Algorithm~1 in the following round. All decisions use the public
optimizer-action history. Small horizons with $m\ge T$ may simply use
Algorithm~1.

We now show that this modification preserves no swap-regret.

%\noindent\textbf{3. Safety and regret.}
Fix any optimizer-action path. On the $\ell$ committed rounds $I$ through
the last passed test, the learners play the fixed distribution $q_j$, and
\begin{equation}
 \left\|\sum_{t\in I}k^{(t)}-\ell(y_j\otimes q_j)\right\|_1
 \le
 \left\|\sum_{t\in I}(e_{c^{(t)}}-z_j)\right\|_1
 +\ell\|z_j-y_j\|_1
 \le m+\ell\varepsilon.
 \label{eq:mc-simple-commitment}
\end{equation}
The initial phase contributes at most $2m$ to cumulative distance from the
target, the triggering round at most $2$, and a fresh suffix of length $R$
contributes at most $C_G\sqrt R$. Since each listed outcome is within $d_T$
of $S_{\NE}$, convexity gives, pathwise,
\[
 \operatorname{dist}_1(\bar k_T,S_{\NE})
 \le d_T+\varepsilon+\frac{3m+2+C_G\sqrt T}{T}=o(1).
\]

Every swap gain at $y_j\otimes q_j$ is at most $2\varepsilon$: a swapped
mixed action cannot outperform the best pure response. 

Since the coefficients
of a swap-gain functional belong to $[-2,2]$,
\eqref{eq:mc-simple-commitment} bounds the commitment contribution by
$2m+4T\varepsilon$. Adding the initial phase, triggering round, and the
fresh Algorithm~1 suffix gives, for every prefix $r\le T$,
\[
 \max_{\phi_i\in\Phi_i}\sum_{t=1}^{r}G_{i,\phi_i}(k^{(t)})
 \le 4T\varepsilon+4m+2+C_{G,i}\sqrt T=:B_i(T)=o(T).
\]

Finally, we show that this modified algorithm makes every Nash-response outcome attainable.

%\noindent\textbf{4. Static attainability and non-manipulability.}
Fix $s\in E_{\NE}$ and choose a listed profile with
$\|s-y_j\otimes q_j\|_1\le\varepsilon$.
Let the optimizer sample independently from $z_j$ in every round.
Writing $d_0=|A_0|$, Hoeffding's inequality and a union bound show that
both decoding and all subsequent tests succeed except with probability
\[
 p_T\le 2\exp\!\left(-\frac{2m}{9M^2}\right)
       +2d_0T\exp\!\left(-\frac{2m^2}{d_0^2T}\right)=o(1).
\]
Here $M=(\log T)^{O_G(1)}$ and $m=\lceil T^{2/3}\rceil$.
On this event, the learners play $q_j$ after the initial phase, giving
$\|\bar k_T-s\|_1\le3m/T+2\varepsilon$.
On its complement the distance is at most $2$. Therefore
\[
 \|\varphi_T(z_j^{\otimes T},\alg'_{\learners})-s\|_1
 \le3m/T+2\varepsilon+2p_T=o(1),
\]
uniformly over $s$. 

%Safety gives the upper menu containment; static attainability and convexity give the reverse containment since $S_{\NE}=\operatorname{conv}(E_{\NE})$.
Fix $\|u_0\|_\infty\le1$. Its maximum $V$ over $S_{\NE}$ is achieved
at some point of the compact generating set $E_{\NE}$.
Upper containment every dynamic optimizer's payoff by $V+o(1)$, and by cooperating with the learners as described above, the optimizer can attain a static payoff of at least $V-o(1)$.
This proves joint non-manipulability.
\end{proof}

\subsection{Proof of Lemma 4.1}
\label{app:optimizer-exploitability}
\begin{proof}
For every $i\in A_1$,
\begin{align*}
    \E_{(a_1,a_2)\sim x\otimes z}
    \left[
        \widehat u_0(R_i,a_1,a_2)
    \right]
    &=
    \E_{a_2\sim z}[A_{i,a_2}]
    -
    \E_{(a_1,a_2)\sim x\otimes z}[A_{a_1,a_2}]\\
    &=
    u_1(i,z)-u_1(x,z).
\end{align*}
Similarly, for every $j\in A_2$,
\[
    \E_{(a_1,a_2)\sim x\otimes z}
    \left[
        \widehat u_0(C_j,a_1,a_2)
    \right]
    =
    u_2(x,j)-u_2(x,z).
\]
The null action gives payoff zero. Hence, maximizing over all optimizer actions
therefore gives exactly
\(
    \operatorname{Exploit}_G(x,z).
\)
\end{proof}

\subsection{Proof of Theorem 4.3}
\label{app:theorem4.3proof}
\begin{proof}
Given a bimatrix game $G$, construct the three-player game $\widehat G$
described above. By Lemma \ref{lem:zero-ce-benchmark},
\(
    V_{\CCE}=0.
\)

Run the purported universal learner algorithms against the following adaptive
optimizer. In round $t$, after observing the learners' current product
distribution
\(
    x_t\otimes z_t,
\)
the optimizer chooses
\[
    c_t
    \in
    \arg\max_{c\in A_0}
    \E_{(a_1,a_2)\sim x_t\otimes z_t}
    \left[
        \widehat u_0(c,a_1,a_2)
    \right].
\]
By Lemma \ref{lem:optimizer-exploitability}, conditional on the history up to this
round and on the learners' current mixed strategies, the optimizer's expected
payoff is exactly
\(
    \operatorname{Exploit}_G(x_t,z_t).
\)

The assumed non-manipulability guarantee therefore implies
\[
    \frac1T
    \sum_{t=1}^T
    \E
    \left[
        \operatorname{Exploit}_G(x_t,z_t)
    \right]
    \le
    \varepsilon.
\]

Now sample
\(
    \tau\sim\operatorname{Unif}([T])
\)
independently of the interaction. Then
\begin{align*}
    \E
    \left[
        \operatorname{Exploit}_G(x_\tau,z_\tau)
    \right]
    &=
    \frac1T
    \sum_{t=1}^T
    \E
    \left[
        \operatorname{Exploit}_G(x_t,z_t)
    \right]\le
    \varepsilon.
\end{align*}

Since exploitability is nonnegative, Markov's inequality gives
\[
    \Pr
    \left[
        \operatorname{Exploit}_G(x_\tau,z_\tau)
        >
        2\varepsilon
    \right]
    \le
    \frac12.
\]
Equivalently,
\[
    \Pr
    \left[
        \operatorname{Exploit}_G(x_\tau,z_\tau)
        \le
        2\varepsilon
    \right]
    \ge
    \frac12.
\]

By definition of $\operatorname{Exploit}_G$, whenever this event occurs,
\[
    u_1(i,z_\tau)-u_1(x_\tau,z_\tau)
    \le2\varepsilon
    \qquad
    \forall i\in A_1
\]
and
\[
    u_2(x_\tau,j)-u_2(x_\tau,z_\tau)
    \le2\varepsilon
    \qquad
    \forall j\in A_2.
\]
Thus
\(
    (x_\tau,z_\tau)
\)
is a $2\varepsilon$-approximate Nash equilibrium of $G$.

The success probability can be amplified by independently repeating the
entire $T$-round experiment. After $r$ independent repetitions, sample one
uniformly random round from each run and compute the exploitability of each
resulting candidate. Return the candidate with minimum exploitability.
Because each run produces a $2\varepsilon$-approximate Nash equilibrium with
probability at least $1/2$, the probability that none of the $r$ candidates
has exploitability at most $2\varepsilon$ is at most
\(
    2^{-r}.
\)

Thus choosing
\(
    r
    =
    \left\lceil
        \log_2(1/\delta)
    \right\rceil
\)
gives success probability at least $1-\delta$.

Finally, the reduction has only polynomial overhead. The constructed
optimizer has
\(
    1+|A_1|+|A_2|
\)
actions. Its payoff table is computable in polynomial time from $A$ and $B$,
and, given $x_t$ and $z_t$, an exploitability-maximizing optimizer action can
be found by evaluating all pure deviations, again in polynomial time.
Likewise, the exploitability of each candidate produced during amplification
is polynomial-time computable.

Therefore, if the learner algorithms run in polynomial time per round and
$T(N,\varepsilon)$ is polynomial in $N$ and $1/\varepsilon$, then for every
inverse-polynomial $\varepsilon$ the above reduction computes a
$2\varepsilon$-approximate Nash equilibrium in randomized polynomial time.

Thus, a polynomial-time universal non-manipulability algorithm with convergence
time polynomial in $N$ and $1/\varepsilon$ would yield a randomized
fully polynomial-time approximation scheme for Nash equilibrium in bimatrix
games. However, \citet{ChenDengTeng2009} show that computing an $\epsilon$-approximate Nash equilibrium for bimatrix games where $\epsilon$ is inverse polynomial in the number of actions of each player is PPAD-complete. Thus, such a randomized polynomial-time approximation scheme would not exist, unless every search problem in PPAD admits a randomized polynomial-time algorithm that, on every input, outputs a valid solution with probability at least $2/3$.

% PPAD-complete and, in particular, that no fully polynomial-time approximation
% scheme exists unless $\mathrm{PPAD}\subseteq\mathrm{P}$. Hence our reduction
% shows that universal multi-learner non-manipulability inherits the
% computational hardness of approximate Nash-equilibrium computation.

\end{proof}

\subsection{Proof of the Joint Approachability Characterization}
\label{app:joint-approachability-proof}

We extend the proof from empirical action distributions to arbitrary vector-valued payoffs $g:A\to\mathbb{R}^d$. As in the preceding sections, the optimizer observes the learners' current product distribution before choosing an action, and the learners observe the optimizer's realized action after each round. We assume that all learners know $g$, and that choices among feasible product strategies and target points use a common, deterministic tie-breaking rule.

Write
\[
    \mathcal Q:=\left\{\bigotimes_{i\in\learners}x_i:
                     x_i\in\Delta(A_i)\right\}
\]
for the compact set of product distributions. For $c\in A_0$ and $q\in\mathcal Q$, define the conditional expected vector payoff
\[
    g(c,q):=\mathbb E_{a_{\learners}\sim q}
                        [g(c,a_{\learners})].
\]

\begin{proof}[Proof of \Cref{thm:joint-approachability-characterization}]
Let $S\subseteq\mathbb{R}^d$ be convex and compact. First suppose $S$ is jointly approachable by $\learners$. We assume, by way of contradiction, that $S$ does not satisfy the joint halfspace condition. Then there exists $U\in\mathbb{R}^d$ such that
\[
    \forall q\in\mathcal Q,\;\exists c\in A_0:\quad
    \langle U,g(c,q)\rangle>h_S(U),
    \qquad h_S(U):=\max_{s\in S}\langle U,s\rangle.
\]
Since $\mathcal Q$ is compact and $q\mapsto \max_{c\in A_0}\langle U,g(c,q)\rangle$ is continuous, there is an $\epsilon>0$ such that
\[
    \forall q\in\mathcal Q,\;\exists c\in A_0:\quad
    \langle U,g(c,q)\rangle\ge h_S(U)+\epsilon.
\]
Fix such a direction $U$. Consider arbitrary learner algorithms $\alg_\learners$. For every round $t$, after observing the learners' product distribution $q^{(t)}\in\mathcal Q$, the optimizer chooses $c^{(t)}\in A_0$ such that
\[
    \langle U,g(c^{(t)},q^{(t)})\rangle\ge h_S(U)+\epsilon.
\]
Conditional on the current product distribution and the optimizer's chosen action, the expected realized vector payoff is $g(c^{(t)},q^{(t)})$. Consequently,
\[
    \mathbb E[\langle U,\bar g_T\rangle]
      \ge h_S(U)+\epsilon.
\]
For every $v\in\mathbb{R}^d$, Cauchy--Schwarz implies
\[
    \operatorname{dist}(v,S)
    \ge\frac{\langle U,v\rangle-h_S(U)}{\|U\|_2}.
\]
Therefore $\mathbb E[\operatorname{dist}(\bar g_T,S)]\ge\epsilon/\|U\|_2$ for every $T$, contradicting joint approachability.

Now suppose $S$ satisfies the joint halfspace condition. We construct a profile of independent learning algorithms in \Cref{alg:joint-approachability-algorithm}. The construction generalizes \Cref{alg:joint-non-manipulable-algorithm}. In round $t$, the learners maintain the cumulative discrepancy
\[
    U^{(t)}:=\sum_{\tau=1}^{t-1}
            \left(k^{(\tau)}-s^{(\tau)}\right)\in\mathbb{R}^d,
    \qquad k^{(\tau)}:=g(c^{(\tau)},q^{(\tau)}),
\]
where each $s^{(\tau)}\in S$. Given $U^{(t)}$, the halfspace condition provides a product distribution $q^{(t)}$ satisfying
\[
    \langle U^{(t)},g(c,q^{(t)})\rangle
       \le h_S(U^{(t)})\qquad\forall c\in A_0.
\]
After observing $c^{(t)}$, the learners choose the same point $s^{(t)}\in S$ such that
\[
    \langle U^{(t)},k^{(t)}-s^{(t)}\rangle\le0.
\]
Such a point exists: any maximizer of $\langle U^{(t)},s\rangle$ over $S$ works. Common tie-breaking allows all learners to maintain the same $U^{(t)}$ using only the public optimizer-action history and their own independently sampled actions.

We have
\begin{align*}
\|U^{(t+1)}\|_2^2-\|U^{(t)}\|_2^2
    &=\|k^{(t)}-s^{(t)}\|_2^2
       +2\langle U^{(t)},k^{(t)}-s^{(t)}\rangle\\
    &\le\|k^{(t)}-s^{(t)}\|_2^2.
\end{align*}
Let $B:=\max_{a\in A,\,s\in S}\|g(a)-s\|_2<\infty$. Since $k^{(t)}$ is a convex combination of vectors $g(c^{(t)},a_{\learners})$, we have $\|k^{(t)}-s^{(t)}\|_2\le B$. As $U^{(1)}=0$, summing over rounds gives $\|U^{(T+1)}\|_2\le B\sqrt T$.

Let $\bar k_T=T^{-1}\sum_{t=1}^T k^{(t)}$ and $\bar s_T=T^{-1}\sum_{t=1}^T s^{(t)}$. By convexity, $\bar s_T\in S$. Thus
\[
    \operatorname{dist}(\bar k_T,S)
    \le\|\bar k_T-\bar s_T\|_2
    =\frac{\|U^{(T+1)}\|_2}{T}
    \le\frac{B}{\sqrt T}.
\]

It remains to pass from conditional expected payoffs $k^{(t)}$ to the realized statistics $g(a^{(t)})$ in the definition of joint approachability. Conditional on the public history, $q^{(t)}$, and the optimizer's realized action $c^{(t)}$, the learners' actions are independently sampled from $q^{(t)}$. Hence
\[
    D^{(t)}:=g(a^{(t)})-k^{(t)}
\]
is a martingale difference sequence. Let $G:=\max_{a\in A}\|g(a)\|_2$. Since $\|D^{(t)}\|_2\le2G$, martingale orthogonality and Jensen's inequality give
\[
    \mathbb E\left\|\frac1T\sum_{t=1}^T D^{(t)}\right\|_2
    \le\frac{2G}{\sqrt T}.
\]
By the triangle inequality,
\[
    \mathbb E[\operatorname{dist}(\bar g_T,S)]
    \le \frac{B+2G}{\sqrt T}=o(1),
\]
as desired. Thus the joint halfspace condition is necessary and sufficient for joint approachability of convex and compact $S$.
\end{proof}

\RestyleAlgo{ruled}
\begin{algorithm2e}[ht]
\caption{Joint approachability algorithm for convex and compact $S\subseteq\mathbb{R}^d$}
\label{alg:joint-approachability-algorithm}
\DontPrintSemicolon
Initialize $U^{(1)}\leftarrow0$\;
\For{$t=1,\ldots,T$}{
    The learners choose a product distribution $q^{(t)}\in\mathcal Q$ satisfying
    $\langle U^{(t)},g(c,q^{(t)})\rangle\le h_S(U^{(t)})$ for every $c\in A_0$\;
    The optimizer observes $q^{(t)}$ and chooses $c^{(t)}\in A_0$\;
    Each learner independently samples her action from her component of $q^{(t)}$\;
    Set $k^{(t)}\leftarrow g(c^{(t)},q^{(t)})$\;
    Choose a common target point $s^{(t)}\in S$ satisfying
    $\langle U^{(t)},k^{(t)}-s^{(t)}\rangle\le0$\;
    Update $U^{(t+1)}\leftarrow U^{(t)}+k^{(t)}-s^{(t)}$\;
}
\end{algorithm2e}

\subsection{Joint response satisfiability and joint approachability are unrelated}
\label{app:JRSdneJA}

\begin{proof}
    We first show there exist convex and compact sets $S\subseteq\mathbb{R}^d$ that are jointly approachable but not jointly response satisfiable.
    
    Consider two learners, each with actions $\{0,1\}$, and let $g(a)=e_{(a_1,a_2)}\in\mathbb{R}^4$ record their joint action; the optimizer's action does not affect $g$. Let
\[
    S=\left\{\tfrac12 e_{(0,0)}+\tfrac12 e_{(1,1)}\right\}.
\]
\newpage
The learners can approach $S$ by alternating deterministically between the joint actions $(0,0)$ and $(1,1)$, without shared randomness. 
%Equivalently, for every direction $U$, at least one of these two pure joint actions has inner product with $U$ no greater than that of their average, so the joint halfspace condition holds. 
However, no single product distribution equals $\tfrac12 e_{(0,0)}+\tfrac12 e_{(1,1)}$, so the joint response-based condition fails. Thus, unlike in the classical two-player setting, response-satisfiability is not necessary for joint approachability.

We now show there also exists convex and compact sets $S\subseteq\mathbb{R}^d$ which are jointly response satisfiable but not jointly approachable.

Consider the game where all three players have actions $\{0,1\}$. Define $$g(c,a_1,a_2):=\mathbf{1}\{c=a_1=a_2\}.$$ Set $u_1=u_2=g$, $u_0=1-g$.

We claim that the set $S=[1/2,1]$ is jointly response satisfiable but not jointly approachable.

$S$ is jointly response satisfiable because for any distribution $y=(q,1-q)$ of the optimizer, the learners can respond by both playing $0$ if $q\geq1/2$ and $1$ otherwise. If $q\geq1/2$, $\mathbb{E}[g(y,0,0)]=q\in S$. If $q<1/2$, $\mathbb{E}[g(y,1,1)]=1-q\in S$.

$S$, however, is not jointly approachable as we show now. Consider any pair of independent learner algorithms. In round $t$, let $p_t:=\mathbb{P}(a_1^{(t)}=0)$ and $r_t:=\mathbb{P}(a_2^{(t)}=0)$. Independence gives $\mathbb{P}(a_1^{(t)}=0,a_2^{(t)}=0)=p_tr_t$ and $\mathbb{P}(a_1^{(t)}=1,a_2^{(t)}=1)=(1-p_t)(1-r_t)$.

At every round, the optimizer can best respond to the distributions the learners commit to, therefore
\begin{align*}
    \mathbb{E}[g(c^{(t)},a_1^{(t)},a_2^{(t)})]&=\min\{p_tr_t,(1-p_t)(1-r_t)\}\\
    &\leq\sqrt{p_tr_t(1-p_t)(1-r_t)}\\
    &=\sqrt{p_t(1-p_t)r_t(1-r_t)}\\
    &\leq\sqrt{(1/4)(1/4)}=1/4\not\in S.
\end{align*}
\end{proof}

% Drop-in section. Requires amsmath, amssymb, and amsthm, with a theorem environment.
% This synthesis incorporates the corrected monitoring argument and the
% expected-realized-regret guarantee for Exp3-based SR_MAB.

\subsection{Unified Algorithm and Guarantee}
\label{sec:unified-algorithm}

For completeness, we conclude the appendix with a unifying theorem that the algorithm as defined in Section \ref{sec:algorithm} when combined with the modifications described in Appendix \ref{app:learner-defection} and $\ref{app:menu-completion}$ has no-swap regret and it jointly non-manipulable.
\newpage
We combine the approachability construction, menu completion, and
learner-defection test into a single horizon-aware profile
$A_{L,T}^{\star}$. Each learner maintains a public virtual protocol and
a private monitoring state. The virtual protocol makes Nash-response
outcomes statically attainable while controlling every optimizer-action
path. The monitoring state protects an individual learner when others
deviate from their prescribed behavior. The learners need not know
$u_0$, and all action samples use independent private randomness.

\paragraph{Model and parameters.}
Fix a finite game with utilities in $[-1,1]$, let
$L=\{1,\ldots,n-1\}$, and suppose $d_0:=|A_0|\ge2$.
Write $\ell=|L|$, $k_i=|A_i|$, and
$\Phi_i=\{\varphi:A_i\to A_i\}$.
The learners know all learner utilities and observe the optimizer's
realized action and their own realized utility after each round.
Other players may adapt to the past and current mixed strategies, but
must choose their current actions without observing an adhering
learner's fresh action sample. All common choices and tie-breaking
rules are deterministic and fixed in advance.

For a known horizon $T$, set
\[
 \varepsilon_T=\frac1{\lceil\log(T+2)\rceil},\qquad
 m_T=\lceil T^{2/3}\rceil,\qquad
 \gamma_T=T^{-1/4},\qquad
 \alpha_T=(T+1)^{-2}.
\]
Also set
\[
 K_\Phi=\sum_{j\in L}|\Phi_j|,\qquad
 \Lambda_T=\log\frac{6TK_\Phi}{\alpha_T},\qquad
 h_i(s)=\left(\frac{k_i}{\gamma_T}+2\right)
                 \sqrt{2s\Lambda_T}.
\]
We describe the construction for $m_T<T$; for the finitely many smaller
horizons, omit menu completion and monitor a fresh approachability run.

\paragraph{Target and public codebook.}
Let
\[
 E_{\mathrm{NE}}
 =\left\{y\otimes q:
       y\in\Delta(A_0),\ q\in\operatorname{NE}(G_L(y))\right\},
 \qquad S_{\mathrm{NE}}=\operatorname{conv}(E_{\mathrm{NE}}).
\]
Take a rational $\varepsilon_T$-net of the full strategy-profile space
under the sum of the players' $\ell_1$ distances. Retain the profiles
$(y_j,(x_{j,i})_{i\in L})$, $j\in[J_T]$, satisfying
\[
 u_i(y_j,b,x_{j,-i})-u_i(y_j,x_j)\le2\varepsilon_T
 \qquad(i\in L,\ b\in A_i),
\]
and write $q_j=\bigotimes_i x_{j,i}$.
For the fixed game, $J_T=(\log T)^{O_G(1)}$.
Fix $c_\star\in A_0$ and set
\[
 M_T=\left\lceil\frac{4J_T}{\varepsilon_T}\right\rceil.
\]
Choose distinct $\theta_j\in\{0,1/M_T,\ldots,1\}$ satisfying
$|\theta_j-y_j(c_\star)|\le\varepsilon_T/2$, and choose
$z_j\in\Delta(A_0)$ with
\[
 z_j(c_\star)=\theta_j,
 \qquad \|z_j-y_j\|_1\le\varepsilon_T.
\]
These choices are possible by the encoding construction in the
menu-completion argument. The entire codebook is common knowledge.

\paragraph{The unified protocol.}
Each learner maintains a copy of the same public virtual state and
initially has private state \emph{monitoring}. The public state starts
in the reading phase, and all monitoring statistics start at zero.
For each round $t$, execute the following steps.
\newpage
\begin{enumerate}
\item \textbf{Compute the virtual prescription.}
The public state determines a product distribution
$\widehat q^{(t)}=\bigotimes_j\widehat x_j^{(t)}$ as follows.
\begin{enumerate}
\item In the \emph{reading phase}, $t\le m_T$, prescribe a fixed pure
learner profile.
\item In the \emph{commitment phase}, prescribe $\widehat q^{(t)}=q_j$
for the decoded label $j$.
\item In the \emph{approachability phase}, maintain $U\in\mathbb R^{|A|}$.
Compute a mixed Nash equilibrium
$(\widetilde y^{(t)},(\widehat x_i^{(t)})_{i\in L})$ of the auxiliary
$n$-player game with the original learner utilities and optimizer utility
\[
 \widetilde u_0(y,x)=\langle U,y\otimes\textstyle\bigotimes_i x_i\rangle.
\]
Set $s^{(t)}=\widetilde y^{(t)}\otimes\widehat q^{(t)}$.
\end{enumerate}

\item \textbf{Choose the actual mixed action.}
If learner $i$ is monitoring, she uses
\[
 \overline p_i^{(t)}
 =(1-\gamma_T)\widehat x_i^{(t)}
       +\gamma_T\operatorname{Unif}(A_i).
\]
Otherwise she uses her private $\mathrm{SR}_{\mathrm{MAB}}$ instance.
The optimizer observes the actual mixed strategies and chooses its
current action. Each adhering learner samples independently from her
actual distribution. After play she observes $c^{(t)}$ and
$r_i^{(t)}=u_i(c^{(t)},a_i^{(t)},a_{-i}^{(t)})$.

\item \textbf{Update the public virtual state using only $c^{(t)}$.}
At the end of round $m_T$, let $\widehat p$ be the observed frequency
of $c_\star$. Decode the unique $j$ satisfying
$|\widehat p-\theta_j|\le1/(3M_T)$, if one exists, and enter commitment.
If decoding fails, enter approachability with $U=0$.
During commitment, enter approachability with $U=0$ at the first failure of
\[
 \left\|\sum_{r=m_T+1}^{t}(e_{c^{(r)}}-z_j)\right\|_1\le m_T.
\]
During an approachability round, update
\[
 U\leftarrow U+e_{c^{(t)}}\otimes\widehat q^{(t)}-s^{(t)}.
\]
All phase changes take effect in the following round. In particular,
$U$ is not updated with a reading or commitment round. Virtual updates
always use $\widehat q^{(t)}$, not the explored or actual distributions.

\item \textbf{Update the private state.}
If learner $i$ was monitoring in round $t$, compute
\[
 \nu_i^{(t)}(b)=u_i(c^{(t)},b,\overline p_{-i}^{(t)}),
 \qquad
 \widehat v_i^{(t)}(b)
 =\frac{r_i^{(t)}\mathbf1\{a_i^{(t)}=b\}}
        {\overline p_i^{(t)}(b)}.
\]
Here $\overline p_{-i}^{(t)}$ always denotes the product of the
\emph{prescribed monitoring distributions}, whether or not other
learners are following them. For every $\varphi\in\Phi_i$, update
\[
 D_{i,\varphi}(t)=D_{i,\varphi}(t-1)
 +\sum_{a\in A_i}\overline p_i^{(t)}(a)
 \left[
 \widehat v_i^{(t)}(\varphi(a))-\widehat v_i^{(t)}(a)
 -\nu_i^{(t)}(\varphi(a))+\nu_i^{(t)}(a)
 \right].
\]
If $\max_\varphi|D_{i,\varphi}(t)|>h_i(t)$, start a fresh
$\mathrm{SR}_{\mathrm{MAB}}$ instance in round $t+1$ and use it permanently.
If already in fallback, update that instance using normalized gain
$(r_i^{(t)}+1)/2$. A private rejection neither resets nor changes the
public virtual protocol.
\end{enumerate}

A public frequency-test failure and a private compliance-test failure
are different events: the former starts the virtual approachability
phase; only the latter sends an individual learner to bandit fallback.
The statistics $D_{i,\varphi}$ can be maintained using the $k_i\times k_i$
matrix from the defection-monitoring construction, without enumerating swaps.

\paragraph{Specification of the bandit fallback.}
For $k_i\ge2$, use Exp3-based $\mathrm{SR}_{\mathrm{MAB}}$ with
\[
 \beta_i=\min\left\{1,\sqrt{\frac{k_i\log k_i}{T}}\right\}.
\]
Initialize row weights $W_{ab}=1$. Each fallback round forms
\[
 Q(a,b)=(1-\beta_i)\frac{W_{ab}}{\sum_cW_{ac}}
                   +\frac{\beta_i}{k_i},\qquad x=xQ,
\]
samples $A\sim x$, observes normalized gain $b$, and updates
\[
 W_{ar}\leftarrow W_{ar}
 \exp\left(
   \frac{\beta_i}{k_i}\,
   \frac{x(a)}{x(r)}b\,\mathbf1\{A=r\}
 \right).
\]
For $k_i=1$ the only action is played. Thus the fallback is specified
using only bandit feedback, and uses the original horizon even after
a random switching time.

\paragraph{Error quantities.}
Write $g_{i,\varphi}(a)
=u_i(a_0,\varphi(a_i),a_{-i})-u_i(a)$ for the coefficient vector of
$G_{i,\varphi}$, and set
\[
 C=\sqrt{2|A|},\qquad
 C_i=\sqrt2\max_{\varphi\in\Phi_i}\|g_{i,\varphi}\|_2,
 \qquad
 d_T=\max_{j\in[J_T]}
         \operatorname{dist}_1(y_j\otimes q_j,E_{\mathrm{NE}}).
\]
Define
\[
 \xi_T=\min\left\{1,
 2\exp\left(-\frac{2m_T}{9M_T^2}\right)
 +2d_0T\exp\left(-\frac{2m_T^2}{d_0^2T}\right)\right\},
\]
\begin{align*}
 a_T&=d_T+\varepsilon_T
                  +\frac{3m_T+2+C\sqrt T}{T},\\
 b_T&=\frac{3m_T}{T}+2\varepsilon_T+2\xi_T,\\
 B_i(T)&=4T\varepsilon_T+4m_T+2+C_i\sqrt T,\\
 e_T&=2\ell\gamma_T+\frac{2\alpha_T}{3}.
\end{align*}
For $k_i\ge2$, let
\[
 F_i(T)=2\left[
 \frac{k_i^2\log k_i}{\beta_i}+2\beta_i T
 +\sqrt{k_iT\left(k_i+\frac{k_i^2}{\beta_i}\right)}
 \right],
\]
and put $F_i(T)=0$ for $k_i=1$.
For every fixed game,
$d_T,\xi_T,a_T,b_T,e_T=o(1)$,
$B_i(T)=o(T)$, and $F_i(T)=O_{k_i}(T^{3/4})$.

\begin{theorem}%[Unified no-swap-regret and non-manipulability guarantee]
\label{thm:unified-guarantee}
The profile $A_{L,T}^{\star}$ above has the following properties.

\begin{enumerate}
\item \textbf{No-Swap Regret.}
For every learner $i$ following the protocol, regardless of the other
learners' admissible behavior and the optimizer's policy,
\begin{align}
 \mathbb E[R_i^{\mathrm{sw}}(T)]
 \le{}&B_i(T)+4\ell\gamma_TT+2h_i(T)
             +4\sqrt{2T\Lambda_T}+2 \notag\\
      &\quad+F_i(T)+\frac{4\alpha_TT}{3}
       =o(T).
 \label{eq:unified-regret}
\end{align}
The maximum over swaps is inside the expectation. In particular,
\[
 \mathbb E[R_i^{\mathrm{sw}}(T)]
 \le B_i(T)+O_G(T^{3/4}\sqrt{\log T}).
\]

\item \textbf{Joint non-manipulability.}
If all learners follow the protocol, then for every
$\|u_0\|_\infty\le1$, writing
$V=\max_{s\in S_{\mathrm{NE}}}\langle u_0,s\rangle$, we have
\begin{align}
 \sup_{\pi^T\in\Pi_T}
 \langle u_0,\phi_T(\pi^T,A_{L,T}^{\star})\rangle
 &\le V+a_T+e_T, \label{eq:unified-dynamic}\\
 \sup_{y\in\Delta(A_0)}
 \langle u_0,\phi_T(y^{\otimes T},A_{L,T}^{\star})\rangle
 &\ge V-b_T-e_T. \label{eq:unified-static}
\end{align}
Consequently, the dynamic-versus-static payoff gap is at most
\[
 a_T+b_T+2e_T
 =a_T+b_T+4\ell\gamma_T+\frac{4\alpha_T}{3}=o(1).
\]

\item \textbf{The completed menu.}
If all learners follow the protocol, define its menu as the closure
of the set of expected CSPs induced by optimizer policies:
\[
 \mathcal M_T^{\star}
 =\overline{\{\phi_T(\pi^T,A_{L,T}^{\star}):\pi^T\in\Pi_T\}}.
\]
Then
\[
 d_{H,1}(\mathcal M_T^{\star},S_{\mathrm{NE}})
 \le\max\{a_T+e_T,b_T+e_T\}=o(1).
\]
\end{enumerate}
All bounds are uniform over the policies indicated, for a fixed finite game.
\end{theorem}

\begin{proof}
\medskip\noindent\textit{The virtual protocol.}
In a fresh approachability run, the auxiliary equilibrium gives
$s^{(t)}\in S_{\mathrm{NE}}$ and
\[
 \langle U,e_c\otimes\widehat q^{(t)}-s^{(t)}\rangle\le0
 \qquad(c\in A_0).
\]
Since both terms are probability distributions,
$\|U_{r+1}\|_2^2\le2r$ after $r$ rounds. Thus every fresh prefix has
cumulative $\ell_1$ distance from $rS_{\mathrm{NE}}$ at most $C\sqrt r$
and mixed-action swap gain at most $C_i\sqrt r$.

Every exact Nash-response profile is within $\varepsilon_T$ of a
retained profile, and compactness implies $d_T\to0$: a subsequential
limit of retained profiles satisfies every exact Nash inequality.
On $r$ commitment rounds through the last passed public test,
\[
 \left\|\sum_t e_{c^{(t)}}\otimes q_j-r(y_j\otimes q_j)\right\|_1
 \le m_T+r\varepsilon_T.
\]
A listed profile has every swap gain at most $2\varepsilon_T$.
Combining this bound with the initial reading phase, the possible
triggering round, and a fresh approachability suffix gives, for every
optimizer-action path,
\[
 \operatorname{dist}_1\left(
 \frac1T\sum_{t=1}^T e_{c^{(t)}}\otimes\widehat q^{(t)},
 S_{\mathrm{NE}}\right)\le a_T
\]
and, for every $s\le T$,
\begin{equation}
 \max_{\varphi\in\Phi_i}\sum_{t=1}^{s}
 G_{i,\varphi}(e_{c^{(t)}}\otimes\widehat q^{(t)})\le B_i(T).
 \label{eq:unified-prefix}
\end{equation}

For $s\in E_{\mathrm{NE}}$, choose a retained $y_j\otimes q_j$
within $\varepsilon_T$ of $s$ and let the optimizer use $z_j$ i.i.d.
throughout the horizon. Hoeffding's inequality and a union bound imply
that decoding and every commitment test succeed except with
probability at most $\xi_T$. On success the virtual average CSP is
within $3m_T/T+2\varepsilon_T$ of $s$, and on failure the distance
is at most $2$. Consequently,
\begin{equation}
 \sup_{s\in E_{\mathrm{NE}}}\inf_{z\in\Delta(A_0)}
 \|\phi_T(z^{\otimes T},A_{L,T}^{\mathrm{virt}})-s\|_1\le b_T.
 \label{eq:unified-attainability}
\end{equation}

\medskip\noindent\textbf{The fallback's expected realized regret.}
For completeness, the conditional expected-realized-regret bound
needed here is $F_i(T)$, rather than only a maximum of expected
fixed-swap gains. Here is a short derivation for $k$ actions and
normalized gains $b_t(\cdot)\in[0,1]$, over $r\le T$ rounds.
Let $p_t$ be the master distribution and $A_t$ its sampled action, and set
\[
 X_{ab,t}=\frac{p_t(a)b_t(b)}{p_t(b)}\mathbf1\{A_t=b\},
 \qquad Y_{ab,t}=\mathbf1\{A_t=a\}b_t(b).
\]
The gain vector may be random but is independent of the fresh sample
conditional on the pre-sampling history. Thus $Y_t-X_t$ is a matrix
martingale difference. Since $p_t(b)\ge\beta/k$,
\[
 \mathbb E\left\|\sum_{t=1}^r(Y_t-X_t)\right\|_F^2
 \le r(k+k^2/\beta).
\]
Cauchy--Schwarz therefore bounds the expected maximum, over all swaps,
of the difference between realized and estimated comparator gains by
$\sqrt{kr(k+k^2/\beta)}$. Applying Blum--Mansour's pathwise
Exp3 inequality (2007, Lemma~10) before maximizing over swaps gives
\[
 \mathbb E[R^{\mathrm{sw}}_{\mathrm{normalized}}(r)]
 \le \frac{k^2\log k}{\beta}+2\beta r
       +\sqrt{kr(k+k^2/\beta)}.
\]
Doubling for utilities in $[-1,1]$ and using $r\le T$ gives $F_i(T)$.
Every step holds conditional on any switching history; hence it
applies to a fresh instance started after the random rejection time.

\medskip\noindent\textbf{No-Swap Regret.}
Fix an adhering learner $i$. While she monitors, let $\mu_{i,\varphi}^{(t)}$
be the actual conditional swap gain, and let
$\overline g_{i,\varphi}^{(t)}
=G_{i,\varphi}(e_{c^{(t)}}\otimes\bigotimes_j\overline p_j^{(t)})$
be the prescribed gain. The importance-weighted estimator is unbiased
for the actual conditional payoff vector, not necessarily for
$\nu_i^{(t)}$. Its centered swap-gain increments are bounded by
$k_i/\gamma_T+2$; realized swap gain minus conditional swap gain has
increments bounded by $4$.

Stop both martingales when $i$ leaves monitoring. Azuma--Hoeffding
and a union bound show that, with probability at least $1-2\alpha_T/3$,
all monitoring prefixes and swaps simultaneously satisfy
\[
 \left|D_{i,\varphi}(s)
 -\sum_{t=1}^s(\mu_{i,\varphi}^{(t)}
                    -\overline g_{i,\varphi}^{(t)})\right|
 \le h_i(s),
\]
and the realized-minus-conditional cumulative gain has absolute value
at most $4\sqrt{2s\Lambda_T}$.
Let $\tau_i$ be the first rejection time and let
$r_i=\min\{T,\tau_i-1\}$, with $r_i=T$ when there is no rejection.
The test has passed through $r_i$, so the absolute cumulative
difference between actual and prescribed conditional swap gains
through that round is at most $2h_i(r_i)$ on this event.

Tensor-product telescoping gives
\[
 \left\|\bigotimes_j\overline p_j^{(t)}-\widehat q^{(t)}\right\|_1
 \le2\ell\gamma_T.
\]
Every swap functional has coefficient norm at most $2$ in $\ell_\infty$.
Using \eqref{eq:unified-prefix}, the pre-trigger realized swap regret
is consequently at most
\[
 B_i(T)+4\ell\gamma_TT+2h_i(T)+4\sqrt{2T\Lambda_T}.
\]
The triggering round contributes at most $2$. The concentration bound
fails with probability at most $2\alpha_T/3$, contributing at most
$4\alpha_TT/3$ in expectation. The conditional fallback bound is
$F_i(T)$. The maximum total gain is at most the sum of the phase-wise
maxima, proving
\eqref{eq:unified-regret}.

\medskip\noindent\textbf{Non-manipulability}
Suppose now that all learners adhere. In the reference execution where
they always use their prescribed monitoring distributions and never
switch, every $D_{i,\varphi}$ is a martingale. The same concentration
and union bound give probability at most $\alpha_T/3$ of any rejection.
Couple actual and reference executions using the same optimizer policy
and random samples until the first rejected test. They agree up to and
including that test, so the actual rejection probability is also at
most $\alpha_T/3$. Their expected CSPs differ in $\ell_1$ distance by
at most $2\alpha_T/3$.

For any adaptive optimizer, compare the reference CSP with the virtual
CSP along the \emph{same realized optimizer-action path}. Exploration
changes the average conditional CSP by at most $2\ell\gamma_T$.
The pathwise virtual containment bound therefore gives
\[
 \sup_{\pi^T}\operatorname{dist}_1
 \bigl(\phi_T(\pi^T,A_{L,T}^{\star}),S_{\mathrm{NE}}\bigr)
 \le a_T+e_T.
\]
Here convexity allows passage from random CSPs to their expectations.
This proves \eqref{eq:unified-dynamic}.

For a static optimizer, couple the virtual and reference executions
using the same i.i.d. optimizer actions. Their virtual prescriptions
then agree throughout. Combining exploration, the rejection coupling,
and \eqref{eq:unified-attainability} gives
\[
 \sup_{s\in E_{\mathrm{NE}}}\inf_{z\in\Delta(A_0)}
 \|\phi_T(z^{\otimes T},A_{L,T}^{\star})-s\|_1\le b_T+e_T.
\]
A maximum of $u_0$ over $S_{\mathrm{NE}}$ is attained in the compact
generating set $E_{\mathrm{NE}}$. This proves
\eqref{eq:unified-static}, and subtraction proves non-manipulability.
Finally, the set of expected attainable CSPs is convex because the
optimizer may randomize initially between policies. Approximating the
generating set $E_{\mathrm{NE}}$ therefore approximates all of
$S_{\mathrm{NE}}$, giving the asserted Hausdorff bound.
\end{proof}

\paragraph{Rates, computation, and scope.}
For the final algorithm the average-regret bound is
\[
 O_G\left(\frac1{\log T}+T^{-1/3}
                       +T^{-1/4}\sqrt{\log T}\right).
\]
The $T^{-1/2}$ approachability rate applies to a fresh approachability
run, not to the completed and monitored algorithm. The non-manipulability
bound also contains $d_T=o(1)$, for which the compactness argument gives
no game-uniform modulus. The construction is horizon-aware and
information-theoretic: its codebook has size $(\log T)^{O_G(1)}$ for
fixed $G$, and its approachability phase computes auxiliary $n$-player
Nash equilibria. No polynomial bound jointly in game size and inverse
accuracy, or anytime conversion preserving static attainability, is
claimed. Private rejection is not an identification of a particular
defector, nor need every deviation be detected; the proof only requires
control of the swap-relevant discrepancies before rejection.

If $|A_0|=1$, omit the codebook and prescribe a fixed Nash equilibrium
of the learner game as the virtual protocol. The same monitoring
argument gives individual no swap regret, and non-manipulability is
automatic because the optimizer has no choice of action. Full menu
equality with $S_{\mathrm{NE}}$ is not asserted in this degenerate case.

\end{document}